\documentclass[aps,prb,superscriptaddress,twocolumn]{revtex4-2}

\usepackage{setspace}
\usepackage{lmodern}

\usepackage[latin9]{inputenc}
\usepackage{amsmath}
\usepackage{amssymb}
\usepackage{graphicx}
\usepackage{epstopdf}
\usepackage{color}
\usepackage{enumerate}
\usepackage{ulem}
\usepackage{bm} 
\usepackage{physics}

\usepackage{array}
\usepackage[labelfont=bf]{caption}
\usepackage{multirow}
\usepackage{floatpag}
\usepackage[misc]{ifsym}
\usepackage{sistyle}
\usepackage{upgreek}
\usepackage[colorlinks=true,linkcolor=blue,citecolor=blue,urlcolor=blue,breaklinks=true]{hyperref}

\begin{document}


\title{Quantum critical collapse of a pinned vortex glass in high-field superconducting resonators}

\author{David Perconte}
\thanks{These two authors contributed equally}
\author{Thibault Charpentier\textsuperscript{*}}
\altaffiliation{Present address: Department of Physics, University of California at Santa Barbara, CA 93106, USA}
\author{Nikolaos Koutsopoulos}
\author{Kalpajit Roy}
\author{Nadjib Benchabane}
\author{Xiaoli Peng}
\author{Florent Blondelle}
\author{Fr\'{e}d\'{e}ric Gay}
\affiliation{Univ. Grenoble Alpes, CNRS, Grenoble INP, Institut N\'{e}el, 38000 Grenoble, France}
\author{Mikhail Feigel'man}
\affiliation{CENN Nanocenter, Ljubljana 1000, Slovenia}
\affiliation{Jozef Stefan Institute, Ljubljana 1000, Slovenia}
\author{Viktor Kabanov}
\affiliation{Jozef Stefan Institute, Ljubljana 1000, Slovenia}
\author{Benjamin Sac\'{e}p\'{e}}
\email{Corresponding author: benjamin.sacepe@neel.cnrs.fr}
\affiliation{Univ. Grenoble Alpes, CNRS, Grenoble INP, Institut N\'{e}el, 38000 Grenoble, France}

\begin{abstract}
\textbf{
The interplay between disorder and vortex-vortex interactions in strongly disordered superconductors in a magnetic field can stabilize a vortex-glass state, characterized by strong pinning and the absence of positional order. Yet its role in the destruction of superconductivity at the field-driven superconductor-insulator transition has remained unresolved. Here we use plasmonic microwave spectroscopy of superconducting resonators patterned from amorphous indium oxide thin films to directly track the superfluid density up to the critical field $B_c$. We find an unexpected resilience of the superfluid density, which decreases only logarithmically over nearly three orders of magnitude in field, in stark contrast to the rapid power-law suppression expected for vortex lattices. We attribute this anomalously slow decay to a collective vortex-pinning mechanism counterintuitively enhanced by vortex-vortex interactions. The superfluid density then vanishes linearly at $B_c$, where independent magnetoresistance measurements identify a continuous quantum critical point, unlike the abrupt transition observed at zero field. We further uncover an exceptionally large nonlinear electromagnetic response of the vortex glass, manifested as a pronounced positive-Kerr effect with potential for quantum sensing. These results establish the pinned vortex glass as an interaction-stabilized superconducting state that preserves superfluid rigidity deep into the mixed phase before undergoing a continuous collapse at the field-driven superconductor-insulator transition.
}
\end{abstract}

\maketitle

Magnetic field and disorder are two fundamental antagonists of superconductivity. In type-II superconductors, a perpendicular magnetic field penetrates the condensate in the form of vortices---topological defects whose motion leads to dissipation---while disorder suppresses phase coherence and can ultimately destroy superconductivity altogether. Their combined action gives rise to a rich variety of vortex states~\cite{Blatter1994} and quantum phase transitions, culminating in the magnetic-field-driven superconductor-insulator transition (SIT)~\cite{Goldman1998,Gantmakher2010,SacepeReview2020}. Despite decades of work, the microscopic mechanisms controlling this transition in strongly disordered superconductors remain poorly understood.

In weakly disordered superconductors, disorder leads to collective pinning of the Abrikosov lattice and to glassy phases such as the Bragg glass, which retains quasi-long-range positional order and admits an elastic description~\cite{Larkin1970,Larkin1979,Blatter1994,GlD1995,Blatter2003}. In contrast, strongly disordered superconductors are characterized by short coherence lengths and suppressed superfluid stiffness~\cite{SacepeReview2020}, where elastic vortex-lattice concepts are expected to fail: vortices are predicted to form a fully amorphous glassy state with proliferating topological defects and a dense spectrum of low-energy excitations~\cite{BraggGlass}. While such a vortex glass has long been anticipated~\cite{MPAFisher1989}, its equilibrium electrodynamic properties and its role in the magnetic-field-driven SIT have been little explored~\cite{sacepe2019}.

DC transport in a magnetic field is strongly influenced by vortex motion and the associated dissipation, and therefore provides only limited information about the underlying superconducting state. Local probes such as scanning tunneling microscopy can instead access the magnetic-field evolution of the superconducting gap and vortex matter~\cite{Ganguly2017, Dutta2022}, revealing field-induced nanoscale inhomogeneity and a loss of global phase coherence while pairing correlations persist. These observations support an important role for phase fluctuations in the field-driven destruction of superconductivity.

In this context, the macroscopic superfluid stiffness, which quantifies the energy cost of a spatial variation of the superconducting phase, is a key quantity for characterizing the transition. Earlier measurements using two-coil mutual inductance~\cite{misra2013}, broadband microwave Corbino spectroscopy~\cite{Liu2011, Liu2013}, and three-dimensional microwave cavities~\cite{crane2007} have established a progressive suppression of the superfluid stiffness upon approaching the field-driven transition. In particular, broadband Corbino measurements on amorphous InO showed the stiffness becoming very small and disappearing close to the transport-defined transition~\cite{Liu2011}. An important question is therefore how the stiffness ultimately vanishes at the quantum critical point, and whether its critical behavior is consistent with the approximately linear suppression suggested by critical-current measurements~\cite{sacepe2019}.

Recent results have, however, revealed unexpectedly different behavior at the zero-field disorder-driven transition. Rather than undergoing the continuous suppression anticipated in conventional scenarios, the superfluid stiffness was found to collapse abruptly, consistent with a first-order quantum transition driven by Coulomb effects~\cite{charpentier2025,Poboiko24}. This unexpected result raises a fundamental question: does the field-driven transition exhibit a similarly abrupt collapse, or does it follow a distinct, continuous route to the destruction of superconductivity? Motivated by this question, here we re-examine the evolution of the superfluid stiffness across the field-driven transition.

\begin{figure*}[ht!]
\centering
\includegraphics[width=0.8\textwidth]{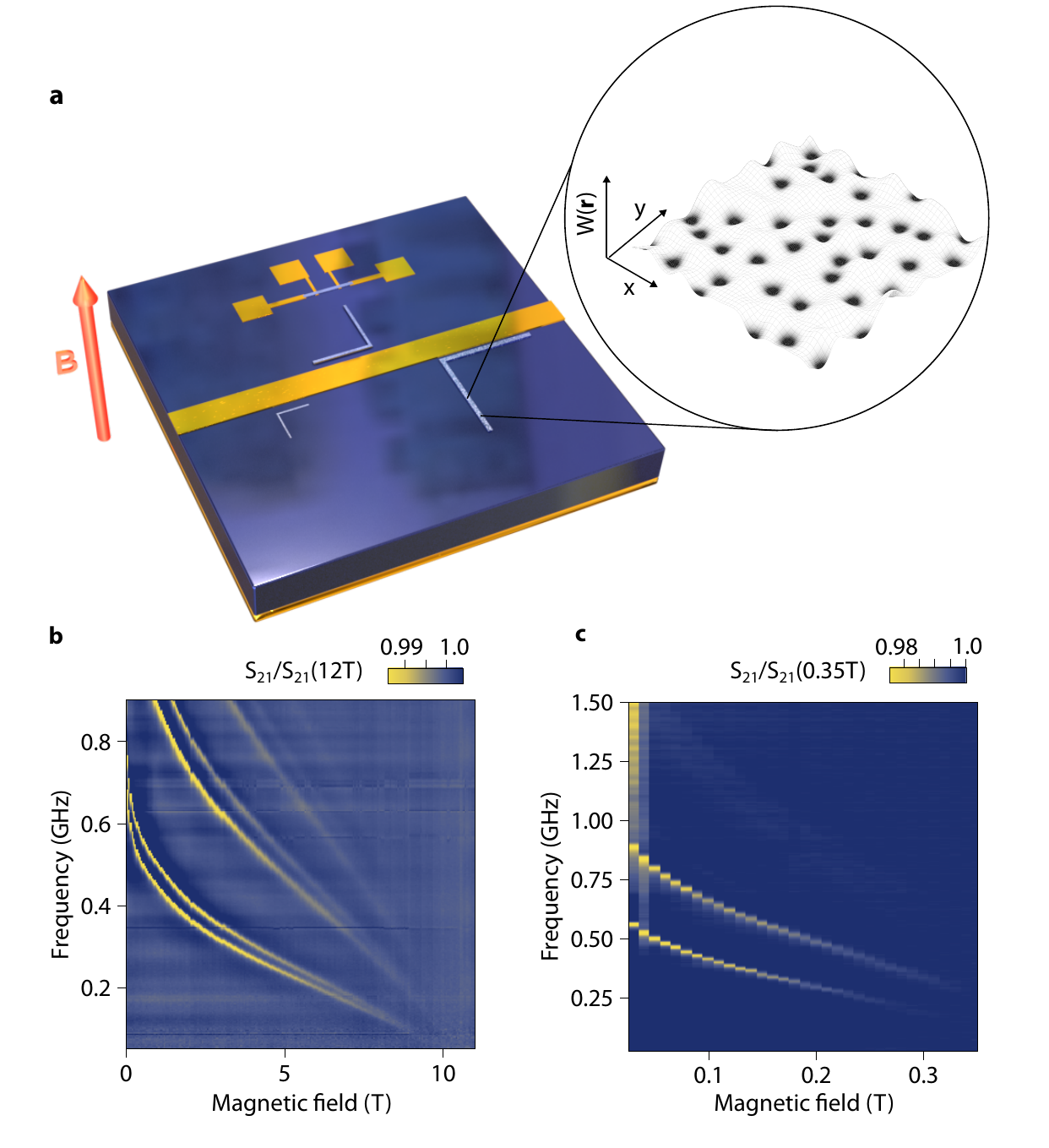}
\caption{\textbf{Superconducting plasmons in the mixed state.}
\textbf{a,} Device schematic. L-shaped a:InO resonators (grey) are capacitively coupled to a $50~\Omega$ impedance-matched gold feedline (center). The magnetic field is applied perpendicular to the substrate. An additional a:InO mesa, contacted by four DC probes, is used for resistance measurements (top). The inset illustrates the random pinning potential $W(\mathbf{r})$ and the resulting vortex glass, represented as black circles.
\textbf{b, c,} Microwave transmission $S_{21}$, normalized to its normal-state value above the critical field, as a function of frequency and magnetic field for resonators DPRes2a and DPRes2b (\textbf{b}), and DPRes11a and DPRes11c (\textbf{c}). In \textbf{b}, higher harmonics corresponding to the two resonators, separated by approximately $50~\mathrm{MHz}$, are also visible. Measurements were performed at $T=0.01~\mathrm{K}$ with microwave powers of $-83~\mathrm{dBm}$ (\textbf{b}) and $-103~\mathrm{dBm}$ (\textbf{c}).
}
\label{Fig1}
\end{figure*}

Here, we used plasmonic microwave spectroscopy of high-kinetic-inductance resonators patterned from amorphous indium oxide (a:InO) films to directly probe the equilibrium superfluid stiffness while remaining insensitive to vortex flow dissipation. We find that, over nearly three orders of magnitude in the magnetic field, the superfluid stiffness exhibits an anomalously slow logarithmic suppression, terminating in a linear vanishing at the critical field, which confirms earlier indirect critical current measurements~\cite{sacepe2019}. We identify this behavior as the hallmark of a new mechanism of collective pinning in an interacting vortex glass, stabilized by the interplay of strong disorder and strong vortex--vortex interactions. In this regime, vortices form a fully amorphous interacting medium in which mutual repulsion generates an effective confining 'cage' ~\cite{cage} for vortex motion, thereby enhancing pinning. 
As a result, the deformation of the vortex glass under superconducting currents is strongly suppressed, leading to a slow field-induced reduction of the macroscopic superfluid stiffness. 
This glassy phase governs the field-induced destruction of superconductivity and leads to a continuous quantum phase transition, in sharp contrast with the abrupt zero-field case.

\subsection*{Interaction-induced collective pinning of glass}

Central to this study is the direct measurement of the equilibrium superfluid stiffness $\Theta(B)$ in the superconducting mixed phase~\cite{misra2013, Liu2013}, a quantity that is inaccessible to conventional transport techniques due to vortex-flow dissipation.
To address this, we extend our prior work~\cite{charpentier2025} by adapting two-tone microwave spectroscopy of superconducting stripline resonators to operate under multi-tesla perpendicular magnetic fields, while simultaneously performing \textit{in situ} dc transport measurements (see Fig.~\ref{Fig1}a and Methods). 

By probing the plasmon dispersion of the resonators at $B=0$ (see Methods), we directly extract the kinetic inductance per square, $L_{\rm K}\sim1-10\,\mathrm{nH/\Box}$ (see Extended Data Table~\ref{tab:SampleParameters}), which governs the two-dimensional superfluid stiffness via $\Theta(0) = (\hbar/2e)^2 / L_{\rm K}$~\cite{charpentier2025}. We emphasize that $\Theta$ characterizes the macroscopic superconducting response of a superconducting current to a variation of the vector potential $\delta \mathbf{A}$, which is an equilibrium thermodynamic quantity existing even in strong magnetic field in case of vortex pinning~\cite{Campbell1,Campbell2,misra2013,Feigelman1987,Feigelman1995,Ocio1997}. Our measurement probes the superfluid stiffness at finite frequencies below 1 GHz rather than the true equilibrium stiffness. However, Suppl. Info. Fig.~S12 shows no detectable frequency dependence, indicating that the measurement is close to the quasistatic limit. 
Upon increasing the magnetic field, we track the frequency shift of the fundamental resonance from its zero-field value $f(0)$ for samples with different levels of disorder (Fig.~\ref{Fig1}b,c). This allows us to directly determine $\Theta(B)/\Theta(0) = \bigl(f(B)/f(0)\bigr)^2$ across the magnetic field--disorder phase diagram, revealing the interplay of disorder and vortex interactions in the superconductor--insulator transition.

\begin{figure*}[ht!]
\centering
\includegraphics[width=0.6\textwidth]{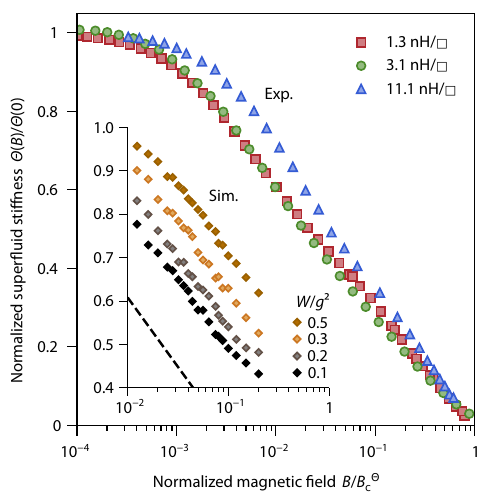}
\caption{\textbf{Interaction-induced collective pinning of the vortex glass.}
Normalized superfluid stiffness $\Theta(B)/\Theta(0)$ as a function of normalized magnetic field $B/B_c^{\Theta}$. Colors correspond to different disorder realizations (characterized by their sheet inductance): DPRes2a (red), DPRes9a (green), and DPRes11a (blue). 
DPRes2a and DPRes11a were measured with microwave powers of $-103~\mathrm{dBm}$ at low field and $-83~\mathrm{dBm}$ at high field, while DPRes9a was measured at $-103~\mathrm{dBm}$ throughout. The corresponding critical fields are $B_c^{\Theta}=9.6~\mathrm{T}$, $5.6~\mathrm{T}$, and $0.44~\mathrm{T}$ for DPRes2a, DPRes9a, and DPRes11a, respectively. 
Inset: Numerical simulations of $\Theta(B)/\Theta(0)$ as a function of normalized vortex density $n/n_c$ (equivalently $B/B_c^{\Theta}$), described in Methods Sec.~\ref{SfB} and SI. The disorder strength $W$ is expressed in units of the vortex interaction constant $g^2 \equiv 2\pi\Theta(0)$. The correlation length is $r_c = 2\xi_0$ for $W/g^2=0.1$, and $r_c = 2.5\xi_0$ for larger disorder strengths. The dashed line in the inset represents the experimental data from the main panel.
}
\label{Fig2}
\end{figure*} 

Figure~\ref{Fig2} presents the main experimental result of this work: the anomalous magnetic-field suppression of the normalized superfluid stiffness $\Theta(B)/\Theta(0)$. For three samples with markedly different disorder and $\Theta(0)$ (Extended Data Table~\ref{tab:SampleParameters}), the superfluid stiffness exhibits a virtually identical logarithmic dependence on $B$ over nearly three orders of magnitude, before vanishing at a critical field denoted $B_c^{\Theta}$.

This logarithmic dependence of $\Theta(B)$ is highly unconventional, as it corresponds to a sublinear growth of the kinetic inductance,
$L_{\mathrm{K}}(B) = (\hbar/2e)^2 / \Theta(B)$.
In contrast, the standard description of a thin superconducting film in the mixed state predicts according to Campbell's penetration depth picture~\cite{Campbell1,Campbell2}:
\begin{equation}
L_{\mathrm{K}}(B) = L_{\mathrm{K}}(0) + \frac{\Phi_0 B}{\tilde{k}} ,
\end{equation}
where $\tilde{k}$ is the curvature of the vortex pinning potential (see Methods and e.g., Ref.~\cite{Willa2016}). 
For isolated vortices, $\tilde{k}$ is constant, while in the 2D collective pinning regime, $\tilde{k} \sim 1/B$, as shown in
Methods, Sec.~\ref{SfB}.2. In result, $\Theta(B)$ is expected to show a crossover between $1/B$ behavior at relatively low fields to faster  $1/B^2$ decay at higher fields, if collective pinning regime sets in.

To highlight the anomalous field dependence in a:InO, we performed a control experiment on a much less disordered superconductor, MoGe, whose kinetic inductance is only $70\,\mathrm{pH/\Box}$, nearly two orders of magnitude smaller. As shown in Extended Data Fig.~\ref{Ext_Fig2}, the MoGe superfluid density exhibits the expected power-law suppression with magnetic field, with an exponent of 1.3, intermediate between the isolated-vortex and collective-pinning regimes.

The logarithmic behavior observed in Fig.~\ref{Fig2} therefore implies an effective pinning parameter that increases with magnetic field as
$\tilde{k}(B) \propto B / \ln(B/B_1)$ where $B_1$ is the penetration field, see Eq.(\ref{equ1}) in Methods, Sec. C. 
We attribute this anomalous behavior to the combined effects of strong vortex--vortex repulsion and strong pinning by local disorder.
In the weak-disorder regime considered in Refs.~\cite{Larkin1970,Larkin1979}, macroscopic pinning becomes progressively weaker with increasing magnetic field because the growing rigidity of the vortex lattice, enhanced by vortex--vortex interactions, reduces the effectiveness of disorder.  As a result, $\tilde{k}$ decreases and the kinetic inductance $L_{\mathrm{K}}(B)$ increases faster than linearly.
In contrast, in the present case the system evolves continuously from a regime of independently pinned vortices to a pinned vortex glass (PVG), without ever forming an elastic vortex lattice (see Methods, Sec.~\ref{LowB} for quantitative estimates). In this regime, vortex--vortex repulsion counterintuitively enhances, rather than weakens, pinning.

As shown in Fig.~\ref{Fig2}, the PVG occupies a broad magnetic-field range above
$B_{\mathrm{cross}} \sim 3 \times 10^{-3} B_c^{\Theta}$.
In this regime, the dominant contribution to the effective pinning parameter $\tilde{k}(B)$ arises from the curvature of the inter-vortex repulsive interaction,
\begin{equation}
\tilde{k} \simeq \beta(B) \left. \frac{\partial^2 U(r)}{\partial r^2} \right|_{r = a_B}
= 2\pi \beta(B)\, \Theta(0)\,\frac{B}{\Phi_0} ,
\end{equation}
where $U(r) = -2\pi\Theta(0)\ln r$ is the standard logarithmic vortex--vortex interaction potential, $a_B = \sqrt{\Phi_0/B}$ is the mean intervortex spacing, while $\beta(B)$ is some slow function of $B$.
The leading dependence $\tilde{k}(B) \propto B$ can be understood within the cage model~\cite{cage}, which assumes nearly uncorrelated displacements of neighboring pinned vortices under an AC drive. A closely related situation arises in the strong-coupling limit of theories of the boson peak in structural glasses~\cite{Boson1,Boson2,Parisi2005}. We show numerically in \textit{Methods}, Sec.~\ref{SfB}, that the spectrum of vortex vibrations around their frozen positions resembles that of structural glasses, providing further support for the relevance of the cage model~\cite{cage} in our system.

\begin{figure*}[ht!]
\includegraphics[width=\textwidth]{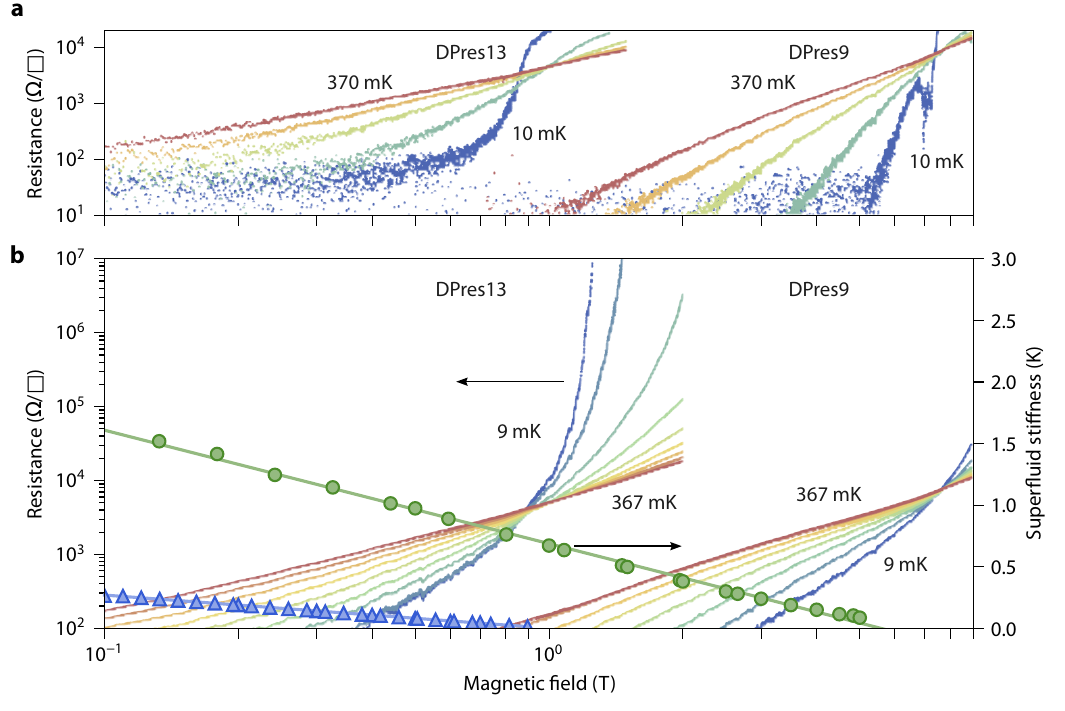}
\centering
\caption{\textbf{Magnetic-field-driven superconductor-insulator transition.}
\textbf{a,} Sheet resistance versus magnetic field measured in a current-bias configuration ($0.1$--$1~\mathrm{nA}$) for samples DPRes9 and DPRes13 at temperatures 10 mK (blue), 90mK, 190mK, 280mK and 370mK (red). The anomalous kink in the $10~\mathrm{mK}$ curve of DPRes9 is reproducible and could originate from low-temperature sensitivity to inhomogeneous current paths~\cite{Seibold12} in the narrow strip of the mesa Hall bar. 
\textbf{b,} Sheet resistance of the same samples measured in voltage bias, together with the corresponding superfluid stiffness (blue triangles and green circles) extracted from the on-chip resonators, as a function of magnetic field for samples DPRes9 and DPRes13 at temperatures 9 (blue), 50, 96, 142, 188, 233, 278, 323, 367 mK (red). The stiffness is obtained from resonators DPRes9c and DPRes13c. Resistance curves were measured in $50~\mathrm{mK}$ steps using bias voltages of $20~\upmu\mathrm{V}$ (DPRes9) and $50~\upmu\mathrm{V}$ (DPRes13). Solid lines are fits of the superfluid stiffness to $\Theta = A \log(B/B_c^{\Theta})$, yielding $A = -0.12~\mathrm{K}$ and $-0.4~\mathrm{K}$, and $B_c^{\Theta} = 5.7~\mathrm{T}$ and $1.0~\mathrm{T}$ for DPRes9c and DPRes13c, respectively.}\label{Fig3}
\end{figure*}

The slower correction $\beta(B) \propto 1/\ln(B)$, which ultimately gives rise to the logarithmic dependence of $\Theta(B)$, is more difficult to capture analytically. We therefore support this interpretation with numerical simulations of logarithmically interacting particles in a random potential (see Methods, Sec.~\ref{SfB} and SI). 
In these simulations, the magnitude of the disorder potential $W$ is estimated by comparing the experimental low-field behavior of $\Delta \Theta(B)$ for $B < 2 \times 10^{-3} B_c$ with numerical results in the corresponding density range $0.01 < n/n_c \leq 0.06$ (see Methods, Sec.~\ref{LowB} and SI). Here $n$ denotes the areal density of logarithmically interacting particles, and $n_c$ is the maximal density corresponding to $B=B_c^{\theta}$. 
This analysis yields $W/2\pi\Theta(0) \approx 1.5 \cdot 10^{-2}$ for the less disordered samples DPRes2 and DPRes9, and $W/2\pi\Theta(0) \approx 3 \cdot 10^{-2}$ for the most disordered sample DPRes11.
The corresponding low-field critical current density is estimated (see Methods, Sec.~\ref{LowB}) as $j_c \approx 4.5 \times 10^2\,\mathrm{A/cm^2}$ for DPRes2, in reasonable agreement with the $B \to 0$ extrapolation of the data in Extended Data Fig.~\ref{Fig5}, where $j_c(B \to 0) \approx (3$--$4) \times 10^2\,\mathrm{A/cm^2}$ for the sample DPRes8 with comparable disorder.
The ratio of the critical current density to the pair-breaking current density,
$j_c / j_0 \sim W/\Theta(0) \gtrsim 0.1$,
is sufficiently large to suppress the conventional collective pinning regime~\cite{Larkin1970,Larkin1979,FeigGeshLarkin1990} and instead favor the formation of a pinned vortex glass.
Unlike previously studied cases~\cite{Zeldov1996,BraggGlass}, the vortex glass in our system emerges directly from the single-vortex pinning regime (see Methods, Sec.~\ref{LowB}).

We found the reduced logarithmic slope $\mathcal{S} = d(\Theta/\Theta_0)/d\ln B$ to be numerically independent of the disorder strength $W$ over the range $W \in (0.6 - 3)\cdot\Theta(0)$, but becomes $W$-dependent at lower disorder.
By contrast, $\mathcal{S}$ depends more sensitively on the correlation length $r_c$ of the random potential (see SI for extended numerical results).
Numerical simulations in the range $0.1 \leq W/g^2 \leq 0.5$, shown in the inset of Fig.~\ref{Fig2} (see Methods, Sec.~\ref{SfB}), indicate that choosing $r_c \sim 2\xi_0$ yields values of $\mathcal{S}$ consistent with the experiment (here $g^2=2\pi\Theta(0)$). 
The remaining vertical offset between numerical and experimental curves reflects the limited access to sufficiently small values of $W/\Theta(0)$ in simulations, which would be required for quantitative alignment~\cite{low-n-problem}. Still, the extrapolation of the numerical results to lower $W$ is consistent with the data (Fig.~\ref{FigS8}). 
Note that we performed a controlled numerical analysis of a model with $1/r$ vortex--vortex repulsion and obtained a power-law suppression of the superfluid density with magnetic field (see SI), supporting the validity of our analytical framework.

A defining feature of the PVG revealed here is that strong vortex--vortex interactions \textit{enhance} pinning efficiency, in stark contrast to the well-established behavior of collectively pinned vortex lattices~\cite{Larkin1979,Blatter1994,BraggGlass}, where the same interactions instead \textit{compete} with pinning by local disorder. This inversion arises from the fundamentally different nature of low-energy excitations in a glassy state. Whereas an elastic vortex lattice supports only long-wavelength shear modes, a vortex glass possesses a dense spectrum of localized deformation modes, providing many additional channels for accommodating disorder-induced distortions. As a result, interactions that stiffen an elastic lattice act to reinforce pinning in the glassy phase.

\subsection*{Quantum critical collapse}

These results raise the question of how the superfluid stiffness of the pinned vortex glass collapses as the system is tuned through the $B$-driven SIT. In amorphous indium oxide, the nature of the state terminating superconductivity depends sensitively on the level of disorder. At low disorder, the transition corresponds to that of a conventional dirty superconductor~\cite{Gantmakher1998,Steiner05,Sacepe15}. Upon approaching the critical disorder, however, the magnetic field, $B_c$, at which superconductivity breaks down decreases and coincides with the emergence of a giant magnetoresistance peak characteristic of a Cooper-pair insulator~\cite{Gantmakher1998,Shahar2004,Steiner05,Shahar2005,ovadia2013,Sacepe15}. In all cases, a crossing point of resistance isotherms is observed and is commonly used as an experimental indicator of the SIT quantum critical point, with critical exponents extracted from finite size-scaling analysis~\cite{Goldman1998,Gantmakher2010,SacepeReview2020}.

Figure~\ref{Fig3} shows the magnetoresistance isotherms for two samples of moderate (DPRes9) and high disorder (DPRes13), together with the corresponding superfluid stiffness $\Theta(B)$. The logarithmic scale enables us to capture the divergence of the resistance over several decades in the vicinity of the transition to the Cooper-pair insulator. 
To accurately probe both the thermal transition into the superconducting state ($B < B_c$) and the transition into the insulating state ($B > B_c$), the measurements were performed in current bias (Fig.~\ref{Fig3}a) and voltage bias (Fig.~\ref{Fig3}b), respectively, using low excitation to minimize nonlinear effects and instabilities near $B_c$~\cite{Doron17}. The current-bias configuration enables reliable access to the superconducting phase, where the resistance drops to the noise floor in the PVG regime for $B < B_c$. In contrast, voltage bias cannot resolve this vanishing resistance but accurately captures the rapid divergence of resistance in the Cooper-pair insulating phase for $B > B_c$.

In Fig.~\ref{Fig3}b, we observe for both samples that the logarithmic suppression of the stiffness naturally approaches a linear vanishing, $\Theta(B) \propto B_c^{\Theta} - B$, as the critical fields $B_c^{\Theta} = 6.0\,\mathrm{T}$ and $0.9\,\mathrm{T}$ are approached for the moderate- and high-disorder samples, respectively. 
The lowest superfluid stiffness accessed in our measurements corresponds to an energy scale of $14\,\mathrm{mK}$, implying an exceptionally large kinetic inductance of $L_{\mathrm{K}} = 560\,\mathrm{nH/\Box}$ for sample DPRes13.

Four consequences follow. First, the direct measurement of a linear suppression of the superfluid stiffness confirms earlier conclusions inferred from critical current measurements performed on similar films. In Ref.~\cite{sacepe2019}, and as reproduced in this work (Extended Data Figure \ref{Fig5}), the critical current density in strongly disordered superconducting films was found to scale as $j_c(B) \sim (B_c - B)^{3/2}$. 
Within a mean-field Ginzburg--Landau framework and assuming strong vortex pinning, this scaling directly implies a linear suppression of the superfluid stiffness, $\Theta(B) \sim (B_c - B)$ ~\cite{sacepe2019}. Such behavior is expected in disordered superconductors from magnetic-field-induced pair breaking. Its robust observation across all our films, however, including those with a strongly developed pseudogap~\cite{sacepe2011}, is unexpected. In the latter case, $B_c^\theta$ remains much smaller than the pair-breaking field, pointing to a more general mechanism underlying this mean-field-like behavior.

Second, although quantum-critical-point (QCP) arguments provide the standard framework for discussing SIT phenomena~\cite{fisher1989,Fisher90}, their applicability is not automatic in the presence of strong disorder and nontrivial effective dimensionality. In our samples, the film thicknesses $d=45$--$70\,\mathrm{nm}$ (see Extended Data Table~\ref{tab:SampleParameters}) exceed the low-temperature correlation length $\xi_c$, which we extract from the measured $B_c$ values using $B_c=\Phi_0/2\pi\xi_c^2$. The resulting $\xi_c$ values lie in the range $5$--$20\,\mathrm{nm}$, with the largest values found in the most strongly disordered films exhibiting a pseudogap. We conjecture that this increase of $\xi_c$ in the pseudogap regime, where phase fluctuations dominate~\cite{sacepe2011,Dubouchet18,charpentier2025}, reflects a correlation length controlled by the superfluid stiffness rather than by the superconducting gap, in contrast to the conventional coherence length. Consequently, the $T=0$ transition cannot \textit{a priori} be regarded as strictly two-dimensional, but instead lies in a mixed 2D--3D regime. Moreover, our previous work~\cite{charpentier2025} showed that the disorder-driven $T=0$ transition is discontinuous (first order), whereas the temperature-driven transition follows a two-dimensional Berezinskii--Kosterlitz--Thouless scenario. Taken together, these observations indicate that the observed linear collapse of the superfluid stiffness occurs in a regime where conventional QCP universality is not assured and must therefore be interpreted with particular care~\cite{Rogachev20}.

Third, the same critical behavior of $\Theta(B)$ is observed across films spanning widely different disorder strengths, with critical fields $B_c$ varying by nearly a factor of $20$. This robustness indicates a common underlying mechanism governing the magnetic-field-driven transition, despite the markedly different magnetoresistance phenomenology in low- and high-disorder films (this work and Refs.~\cite{Gantmakher1998,Steiner05,Sambandamurthy2006,Sacepe15}).

An important open question concerns how the discontinuous disorder-driven transition at $B=0$~\cite{charpentier2025} connects to the apparently continuous magnetic-field-driven transition. Notably, the most disordered film studied here still exhibits a zero-field stiffness $\Theta(0)$ that is by factor 1.5 above the critical value $\Theta_{\mathrm{min}}(0)$ identified in Ref.~\cite{charpentier2025}. This observation suggests that the magnitude of the discontinuity in the field-driven transition, if any, diminishes rapidly as disorder is reduced.

Fourth, while magnetoresistance measurements alone may sometimes suggest anomalous metallic or failed superconducting behavior below $B_c$~\cite{Breznay17,kapitulnik2019,Tamir19} (see, for example, Fig.~\ref{Fig3}b), the direct observation of a finite superfluid stiffness that vanishes continuously at the transition demonstrates that superconductivity persists up to the critical field and collapses only at $B_c$. This finding supports earlier demonstrations that the anomalous metallic behavior arises from the extreme sensitivity of the mixed phase to external noise~\cite{Tamir19}.

\subsection*{Positive Kerr nonlinearities and dissipation}

\begin{figure*}[ht!]
\includegraphics[width=0.7\textwidth]{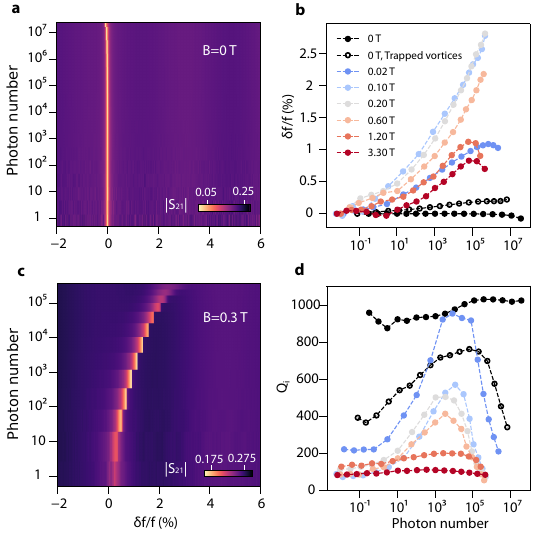}
\centering
\caption{\textbf{Positive Kerr nonlinearity and dissipation.}
\textbf{a, c,} Microwave transmission $S_{21}$, in arbitrary units, as a function of photon number and frequency shift $\delta f/f$ relative to the low-photon-number resonance frequency, at $B = 0~\mathrm{T}$ (\textbf{a}) and $B = 0.3~\mathrm{T}$ (\textbf{c}), for sample DPRes8.
\textbf{b,} Relative frequency shift $\delta f/f$ versus photon number, extracted at various magnetic fields.
\textbf{d,} Internal quality factor $Q_{\mathrm{i}}$ versus photon number, at various magnetic fields; the legend is the same as in \textbf{b}.
In \textbf{b} and \textbf{d}, filled black circles correspond to measurements performed after zero-field cooling, whereas the open black circle corresponds to a measurement with trapped vortices. 
Measurements were performed at $T=0.01~\mathrm{K}$.
}
\label{Fig4}
\end{figure*}

Another way to probe the stiffness and pinning of the vortex glass is through its nonlinear response under an AC drive. Na\"{\i}vely, shaking vortices via the Magnus force would be expected to weaken pinning and reduce the superfluid stiffness. Instead, as shown in Fig.~\ref{Fig4}c, we observe the opposite behavior: the stiffness increases with microwave power, leading to a positive frequency shift as a function of intra-cavity photon number.

The magnitude of this effect, reaching up to 2.5\%, is remarkably large and contrasts sharply with the essentially linear response observed at $B=0$ (Fig.~\ref{Fig4}a), where no vortices are present. As shown in Fig.~\ref{Fig4}b, such a positive Kerr nonlinearity indeed emerges once vortices enter the resonator, culminates at $B=0.3$ T, and decreases upon further increase of magnetic field. 

Strikingly, the resonator dissipation decreases with increasing microwave power. Figure~\ref{Fig4}d displays the photon number-dependent internal quality factor, $Q_{\rm i}$, for the same magnetic fields indicated in the legend of Fig.~\ref{Fig4}b. 
At $B=0$~T, $Q_{\rm i}$ increases monotonically with power~\cite{charpentier2025}, most likely due to conductive losses~\cite{charpentier2025,charpentier2025b} or the saturation of localized electronic degrees of freedom~\cite{Khvalyuk2025}. Once vortices enter the resonator, however, the dissipation becomes non-monotonic: $Q_{\rm i}$ increases by up to a factor of three before decreasing again at the highest powers.

These observations can be understood in terms of nonlinear vortex-pinning effects arising from the out-of-equilibrium population of Caroli--de Gennes--Matricon vortex-core states, that is, vortex-core overheating. As discussed in Methods~\ref{power-dep-Q}--\ref{power-stiffness}, at frequency $\omega \approx 1$~GHz and low temperature $T \approx 10$~mK, quasiparticles localized within a vortex core undergo inelastic scattering with relaxation time $\tau_{\text{in}} \gg 1/\omega$, leading to an effective increase of their temperature under microwave drive. This power-dependent overheating enhances the effective spring constant $\tilde{k}$~\cite{PAF2023}, i.e. strengthens vortex pinning, and thereby increases the imaginary part of the complex conductivity. This mechanism accounts for the substantial and counterintuitive positive Kerr effect, which translates into a power-induced enhancement of the superfluid stiffness observed in Fig.~\ref{Fig4}b,c.

Regarding the dissipation suppression with power, a theoretical description must account for the discreteness of vortex-core states that results from the unusually large mean level spacing of electronic states inside a vortex core. In our strongly disordered superconductor, with coherence length $\xi_0 = 5$~nm and Fermi wavevector $k_F \approx 4$~nm$^{-1}$~\cite{Sacepe15}, the level spacing can reach $\hbar\omega_0 \sim \Delta/(k_F\xi_0) \sim 10\,\mu\mathrm{eV}$, substantially exceeding the microwave photon energy $\hbar\omega$. Under these conditions, the discrete nature of vortex-core states becomes essential and the continuum approximation underlying the conventional Kubo formalism breaks down, see Methods~\ref{power-dep-Q} for a more detailed discussion.

A vortex core can therefore be viewed as a isolated, driven quantum system, particularly at low magnetic field where vortices are well separated, undergoing dynamical localization~\cite{Basko2003,Silva2020}, a mechanism in which energy absorption and average dissipation are suppressed. Upon increasing magnetic field, the inter-vortex distance decreases, enabling hybridization between electronic states in neighboring vortex cores. Dissipation then increases, and the response progressively approaches the behavior expected from the conventional Kubo description.

Interestingly, the robustness of the resonators at multi-tesla fields and their nonlinearities suggest a route toward circuit quantum electrodynamics (QED) sensing of a wide range of magnetic-field-induced quantum systems, including spin qubits~\cite{Burkard23}, quantum Hall devices~\cite{Appugliese22,enkner2025tunable} hosting localized (fractional) charges~\cite{diLuca2025} or magnonic excitations~\cite{assouline2021excitonic}, and quantum Hall superconducting hybrids~\cite{Vignaud23}. 
This prospect is further strengthened by the substantial increase in the resonators' characteristic impedance, $Z(B)=\sqrt{L_K(B)/C}$, arising from the divergence of the inductance (see Extended Data Fig.~\ref{Ext_Fig5} and Methods \ref{method:strong_coupling}). The resulting enhancement of the zero-point voltage fluctuations promotes strong coupling to electric dipole moments of quantum systems~\cite{Clerk2020} and may even enable access to the ultra-strong coupling regime~\cite{FornDiaz2019,FriskKockum2019}, opening new opportunities for high-field QED experiments.

\subsection*{Discussion}

The pinned vortex glass and the suppression of the superfluid stiffness uncovered here challenge the conventional understanding of magnetic-field-driven superconductor--insulator transitions in strongly disordered systems. In particular, it raises the fundamental question of how a continuous field-tuned transition can coexist with a first-order disorder-driven transition at $B=0$~\cite{charpentier2025}, and whether these two instabilities belong to a unified phase diagram or reflect distinct underlying mechanisms. The sharp suppression of the critical field $B_c$ with increasing disorder~\cite{Sambandamurthy2006}, well below the pair-breaking scale~\cite{Sacepe15}, further indicates that the controlling energy scale is not captured by standard mean-field or collective-pinning approaches.

The observed logarithmic $\Theta(B)$ dependence is qualitatively inconsistent with both independent pinning of vortices by disorder and collective pinning of a deformed vortex lattice. This suggests a new scenario: the pinning of an interacting vortex glass. We support these observations by detailed numerical analysis, while an analytical theory of such a vortex state is currently under development.

Our observations also point to unresolved microscopic issues. The unconventional dissipation and dynamical-localization effects inferred in vortex cores call for a quantitative description of nonequilibrium quasiparticle dynamics in the presence of strong disorder and discrete level structure, including the temperature dependence of the inelastic scattering time. More broadly, the nature of the insulating state emerging above $B_c$~\cite{Gantmakher1998,Shahar2004,Steiner05,Shahar2005,ovadia2013,Sacepe15}, particularly when pairing survives far beyond the transition, remains to be clarified and may be connected to reported finite-temperature insulating behavior~\cite{Ovadia15}.

These open questions motivate the development of an analytic theory of the pinned vortex glass beyond conventional approximations. Such a framework may not only resolve the magnetic-field-driven transition in disordered superconductors but also illuminate the broader physics of interacting quantum glasses, including Coulomb-glass systems~\cite{MuellerIoffe,MuellerPankov}

\section*{Methods}

\subsection{Sample fabrication}

Our a:InO and MoGe resonators were patterned by electron-beam lithography using polymethyl methacrylate (PMMA) resist, cold development, and liftoff. a:InO films with thicknesses of 44--70~nm (see Extended Data Table~\ref{tab:SampleParameters}) were deposited by electron-beam evaporation of high-purity (99.999\%) In$_2$O$_3$ onto high-resistivity silicon substrates. The disorder was tuned by controlling the O$_2$ partial pressure during deposition and the film thickness. MoGe films were deposited by DC magnetron sputtering from a Mo$_{0.75}$Ge$_{0.25}$ target. 

The backside of each chip was coated with a thick gold layer serving as a ground plane for the microstrip resonators. Each chip hosts two to three resonators with varying geometries (width and length; see Extended Data Table~\ref{tab:SampleParameters}), capacitively coupled to a gold feedline. Electromagnetic simulations were performed to optimize the resonator coupling ($Q_i \approx Q_c$) in regimes where internal losses are significant, such as at high magnetic fields. Transport measurements were carried out on co-deposited Hall-bar structures on the same chip, with dimensions listed in Extended Data Table~\ref{tab:SampleParameters}, enabling direct comparison between magnetotransport and microwave response.

\subsection{Measurements}

Microwave measurements are performed in transmission using a vector network analyzer. The transmitted signal is amplified by a cryogenic low-noise amplifier (10--2000~MHz bandwidth) developed at Arizona State University. The amplifier input is filtered by a 1-m-long dissipative coaxial cable wound around a copper tube and thermalized with silver epoxy.

The resonators are designed such that the $B=0$~T resonances fall within the 1--2~GHz range and subsequently shift to lower frequencies under magnetic field while remaining within the amplifier bandwidth.

Resonances appear in the transmission amplitude at frequencies
\[
f_n = \frac{n}{2L\sqrt{l\,c_n}},
\]
for mode number $n$, where $l = L_K^{\square}/w$ is the kinetic inductance per unit length and $c_n$ is the capacitance per unit length. The exceptionally high zero-field kinetic inductance, $L_K^{\square}(0) \sim 1$--$8$~nH/$\square$, shifts the resonant modes to low frequencies (below 1~GHz). Upon increasing magnetic field, $L_K(B)$ increases, resulting in a decrease of $f(B)$, from which we extract the relative field-dependent superfluid stiffness,
\[
\frac{\Theta(B)}{\Theta(0)} = \left(\frac{f(B)}{f(0)}\right)^2.
\]
The zero-field kinetic inductance is determined by matching the measured zero-field resonance frequency with electromagnetic simulations of the same geometry with varied input $L_K$. When accessible, the obtained value is further confirmed by using two-tone spectroscopy, as described in Ref.~\cite{charpentier2025} for similar samples.

Magnetotransport measurements on co-deposited Hall-bar structures are carried out using either current or voltage bias to accommodate the large resistance variations between the superconducting and insulating states. AC excitation currents of order $0.1$--$1$~nA and voltages of $10\,\mu\mathrm{V}$ are applied. The critical current $I_c(B)$ is obtained by identifying $I_c$ at the onset of the resistive state from $dV/dI$ versus DC current.

\subsection{Signatures of vortex penetration with magnetic field}
\label{sec:vortex_entry}

The entry of a vortex into a superconductor results in both a local suppression of the superfluid density at its core and a Lorentz force arising from the screening currents circulating around it. The energy trade-off between these contributions makes vortex penetration energetically favorable above a characteristic magnetic field $B_1$~\cite{Likharev1972,Martinis2004}:
\begin{equation}
    \label{equ1}
    B_1 = \frac{2}{\pi}\frac{\Phi_0}{w^2}\ln\!\left(\frac{w}
    {4\xi}\right), 
\end{equation}
where $\Phi_0 = h/2e$ is the superconducting flux quantum, $w$ is the strip width, and $\xi$ is the superconducting coherence length. This onset of vortex penetration is observed experimentally, as shown in Extended Data Fig.~\ref{Ext_Fig4}.

In addition, vortex entry into a superconducting film is expected to be hindered by the Bean--Livingston surface barrier~\cite{Bean1964}, implying that vortices should not penetrate until a higher critical field 
\[
B_2 = \frac{\Phi_0}{2\pi \xi w}.
\]
This effect is not resolved in our measurements, likely due to edge roughness of the superconducting meander, which creates weak spots where vortices nucleate in the range $B_1 < B < B_2$. For a representative resonator width $w = 1~\upmu\mathrm{m}$ and coherence length $\xi = 5~\mathrm{nm}$, the expected critical field is $B_2 \approx 66~\mathrm{mT}$. In contrast, we observe a sharp decrease of the quality factor at magnetic fields of only a few millitesla.

\subsection{Power dependence of quality factor}
\label{power-dep-Q}

The quality factor $Q$ of a resonator in the presence of a magnetic field is determined by dissipation associated with vortices that penetrate the superconducting film once the magnetic field exceeds $B_1$. This effect is visible in Extended Data Fig.~\ref{Ext_Fig4}, where a sharp drop of $Q$ occurs at fields of order $5\,\mathrm{mT}$. In this low-field regime, the variation of the effective superfluid stiffness $\Theta(B)$ remains moderate. Consequently, the dependence of the quality factor,
\begin{equation}
Q = \sigma_2 / \sigma_1 ,
\end{equation}
on the microwave power is primarily governed by the behavior of the real part of the complex conductivity $\sigma_1(\omega)$.

The complex conductivity of a superconductor in the mixed state with pinned vortices can be written as (see, e.g., Refs.~\cite{PAF2023,test-note-error})
\begin{equation}
\sigma(\omega) = \frac{i c^2}{4\pi\omega}
\frac{1}{\lambda_L^2 + \frac{\Phi_0 B d}{4\pi k(\omega)}} ,
\label{sigma1}
\end{equation}
where $k(\omega) = \tilde{k}(\omega) - i\omega\eta$. The frequency-dependent effective spring constant $\tilde{k}(\omega)$ is given by
\begin{equation}
\tilde{k}(\omega) = k(0) - \frac{i\omega\tau_{in}}{1 - i\omega\tau_{in}} \, \delta k .
\label{k1}
\end{equation}

Here $c$ is the speed of light (we use the CGS system of units), $\lambda_L$ is the London penetration depth, $d$ is the film thickness, $\tau_{in}$ is the inelastic relaxation time, and $\eta$ is the vortex friction coefficient, which determines the friction force $f_v = \eta v$ acting on a vortex moving with velocity $v$. The friction coefficient $\eta$ is related to the linear \textit{dc} conductivity in the flux-flow regime through $\sigma_{ff} = c^2 \eta / (B \Phi_0)$. The parameter $\delta k > 0$ represents the difference between the high-frequency and zero-frequency limits of the effective spring constant~\cite{PAF2023}.

At microwave frequencies in the GHz range and at very low temperatures ($T \sim 20\,\mathrm{mK}$), we expect $\omega \tau_{in} \gg 1$. In this limit, $\tilde{k}(\omega) \simeq k_\infty = k(0) + \delta k$, while the imaginary part of $\tilde{k}(\omega)$ is small. Expanding Eq.~(\ref{sigma1}) to leading order in the magnetic field $B$, the dominant contribution to the real part of the conductivity reads
\begin{equation}
\sigma_1 = \eta \, \frac{c^2 \Phi_0 B d}{16\pi^2 \lambda_L^4 \tilde{k}^2} .
\label{sigma2}
\end{equation}

At higher microwave power, nonlinear effects develop, leading to a reduction of the vortex friction coefficient $\eta$ and consequently to a decrease of dissipation. The reduction of vortex friction $\eta$ is due to the discrete nature of localized quasiparticle levels inside vortex core. The effect of level discreteness upon energy dissipation in quantum dots was discussed both analytically~\cite{Basko2003} and numerically~\cite{Silva2020}; its essence is the same as for the well-known saturation effect for dissipation in driven two-level systems.
This mechanism explains the observed increase of $Q$ with microwave power at low magnetic fields.

\subsection{Power dependence of superfluid stiffness}
\label{power-stiffness}

At low magnetic fields, the suppression of the superfluid stiffness can be described in terms of an effective penetration depth $\lambda_{\mathrm{eff}}$. This quantity can be derived from Eq.~(\ref{sigma1}) by expanding the imaginary part of the conductivity, $\Im \sigma$, to leading order in the magnetic field $B$, yielding
\begin{equation}
\lambda_{\mathrm{eff}}^2 = \lambda_L^2 + \frac{\Phi_0 B d}{4\pi \tilde{k}} .
\label{lambda2}
\end{equation}
We note that this equation was written in Ref.~\cite{PAF2023} with an incorrect coefficient $8\pi$ instead of $4\pi$, due to a calculation error.

Using the standard relations for the kinetic inductance of a thin film, $L_K = 4\pi \lambda^2 / d$, and for the superfluid stiffness, $\Theta = (\Phi_0 / 2\pi)^2 / L_K$, we obtain the vortex-induced correction to the superfluid stiffness
\begin{equation}
\frac{\Theta(B)}{\Theta(0)} = 1 - \frac{\Phi_0 B d}{4\pi \lambda_L^2 \tilde{k}} .
\label{Theta-non}
\end{equation}

As discussed in the previous subsection, the effective spring constant $\tilde{k} = k(0) + \delta k$ contains a positive non-equilibrium contribution $\delta k$~\cite{PAF2023}, whose magnitude scales as $T_{\mathrm{eff}}^2$. The effective temperature $T_{\mathrm{eff}}$ of quasiparticles inside a vortex increases with microwave power, leading to an enhancement of $\tilde{k}$ and, consequently, to a positive variation of $\Theta(B)$ with power.

It is important to verify that the \textit{ac} current density in the resonator always remains below its critical value, ensuring that vortices remain pinned even at high photon numbers $N_{\mathrm{ph}}$. The mean-square two-dimensional current density is related to $N_{\mathrm{ph}}$ as (see Appendix~F of Ref.~\cite{khvalyuk2024})
\begin{equation}
\overline{j_{2D}^2} = \left( \frac{2e}{\hbar} \right)^2
\frac{\Theta \, \hbar \omega \, (N_{\mathrm{ph}} + 1/2)}{w l} ,
\label{mean-square-current}
\end{equation}
where $w$ and $l$ denote the width and length of the resonator, respectively.

Using the parameters of sample DPRes2a at a frequency of $0.735\,\mathrm{GHz}$, we estimate the root-mean-square current density as
\begin{equation}
j(N_{\mathrm{ph}}) \approx 0.3 \sqrt{N_{\mathrm{ph}}} \, \frac{\mathrm{A}}{\mathrm{cm}^2}
\leq j_{\mathrm{max}} = 10^3 \, \frac{\mathrm{A}}{\mathrm{cm}^2} ,
\label{jN}
\end{equation}
where the upper bound corresponds to $N_{\mathrm{ph}} \sim 10^7$. Importantly, $j_{\mathrm{max}}$ always remains below the critical current density $j_c$, estimated in Subsection~\ref{LowB}.

\subsection{Superfluid stiffness scaling with magnetic field}
\label{SfB}

At low magnetic fields, vortices are independently pinned by a random potential of typical amplitude $W$. Although the thickness of our films ($d = 45$--$70$~nm) exceeds the low-temperature coherence length $\xi_0$ by roughly an order of magnitude~\cite{Sacepe15}, the vortices can nevertheless be treated as effectively two-dimensional pancake objects. Indeed, the disorder is not sufficiently strong to induce significant bending of vortex lines along their length $d$. The corresponding estimates made along the lines of the approach discussed in particular in Ref.~\cite{Blatter1994}
demonstrate that the transverse displacement of vortex lines, $u_\perp(d)$, remains bounded by the pinning correlation length $r_c \approx 2\xi_0$, which is always smaller than the inter-vortex spacing $a_B = \sqrt{\Phi_0/B}$. This hierarchy of length scales allows us to model the system as a two-dimensional assembly of point-like vortices pinned by a two-dimensional random potential. 
In other words, corrections to the interaction energy of slightly curved segments of vortex lines with respect to the same energy within the approximation of straight segments are very small and can be neglected.

\subsubsection{Vortex glass regime}

Upon increasing magnetic field, the interaction between vortices, $
U(r) = -2\pi \Theta(0)\,\ln(r)$, strengthens and begins to compete with individual pinning by the random potential. It is convenient to describe the system of interacting vortices in the presence of disorder through a formal analogy with two-dimensional electrostatics of point particles with effective charge $g=\sqrt{2\pi\Theta(0)}$.

Within this analogy, the renormalization of the macroscopic superfluid stiffness $\Theta(B)$ can be expressed in terms of an effective ``dielectric function'' $\epsilon_v$, in close correspondence with electrostatics in polarizable dielectrics. To this end, we introduce formal two-dimensional analogues of the electric field $\mathbf{E}_v$ and electric displacement $\mathbf{D}_v = \epsilon_v \mathbf{E}_v$, where $\epsilon_v = 1 + 2\pi \chi_v$,(the coefficient is $2\pi$, rather than the usual $4\pi$, because the problem is two-dimensional) and $\chi_v$ is the vortex polarizability, $\chi_v = \langle \partial \mathbf{P}/\partial \mathbf{E} \rangle$ with polarization vector $\mathbf{P} = \frac{g}{\mathcal{A}} \sum_i \mathbf{u}_i$. Here $\mathcal{A}$ is the system area and $\mathbf{u}_i$ is the displacement of the $i$-th vortex induced by a weak applied field $\mathbf{E}$ coupled through the term $-g\,\mathbf{u}_i \cdot \mathbf{E}$ in the Hamiltonian.

In the limit of low magnetic field and independently pinned vortices, one finds $\chi_v = g^2 n/\tilde{k}$, in direct correspondence with Eqs.~(\ref{lambda2},\ref{Theta-non}). When vortices interact both with the random potential and with each other, $\chi_v$ and $\epsilon_v$ become spatially fluctuating quantities, and the appropriate averaging procedure becomes nontrivial. The quantity of primary interest,
\begin{equation}
\theta(B) \equiv \frac{\Theta(B)}{\Theta(0)} = \left\langle \frac{1}{\epsilon_v} \right\rangle = \left\langle \frac{1}{1 + 2\pi \chi_v} \right\rangle,
\label{Theta-full}
\end{equation}
cannot in general be reduced to the inverse of the averaged polarizability if $\chi_v$ is not small.

It is therefore necessary to evaluate Eq.~(\ref{Theta-full}) numerically over a broad range of magnetic fields. We perform simulations of two-dimensional Coulomb particles interacting with a local random potential of variance $W$ and correlation length $r_c$, defined in units of the lattice spacing $a$ of the numerical grid (see Supplementary Information for details).

We find a logarithmic dependence,
\[
\left\langle \epsilon_v^{-1} \right\rangle = A - C \ln(n/n_c),
\]
over a broad range of particle densities $0.01 < n/n_c < 0.2$, where $n_c = 1/a^2$. At higher densities ($n > 0.2\,n_c$), the London approximation for vortex interactions is no longer valid due to suppression of the order parameter by magnetic-field-induced pair breaking.

This logarithmic behavior is observed for various values of the correlation length $r_c$. The parameter $C$ is primarily controlled by the ratio $r_c/\xi_0$, with only weak dependence on $W/g^2$ for $W/g^2 < 0.15$ (see Supplementary Information). We also verified that simulations performed on $50 \times 50$ and $100 \times 100$ lattices yield nearly identical results for $W/g^2 \in (0.1, 0.5)$, indicating negligible finite-size effects in this regime.

To relate the numerical parameter $n/n_c$ to the physical ratio $B/B_{c2}$, we computed the disorder-free magnetization curve $n(B)$ (see Supplementary Information, Sec.~1B). We find that the maximal vortex density $n_c = 1/a^2$ corresponds to the upper critical field $B_{c2} = \Phi_0 / 2\pi \xi_0^2$. This yields the relation between the lattice spacing and coherence length, $a \approx \sqrt{2\pi}\,\xi_0 \approx 2.5\,\xi_0$.

The main numerical results are shown in the inset of Fig.~\ref{Fig2} for $W/g^2 = 0.1, 0.2, 0.3,$ and $0.5$. We use a correlation radius $r_c = 2\xi_0$ for $W/g^2 = 0.1$, and $r_c = 2.5\xi_0$ for larger disorder strengths (see Supplementary Information for details). 

\subsubsection{Spectrum density of the Hessian}

Additional evidence for the nature of the pinned vortex glass is provided by the analysis of the spectrum of the Hessian matrix
$\hat{H} = \frac{\partial^2 E{\mathbf{r}_i}}{\partial r_i^a\partial r_j^b}|_{\textrm{min}} $
where $E_{\mathbf{r}_i}$ is the energy functional of logarithmically interacting 2D particles in a random potential, and the matrix $\hat{H}$ is evaluated at its minimum. We computed the Hessian $\hat{H}$ numerically (see Supplementary Information IV for details) and diagonalized it to find the density of its eigenvalues $\nu(\lambda)$, which is an analogue of the vibration density of states $\rho(\omega^2)$ in model glasses~\cite{Boson1,Boson2}, typically showing a "boson peak" in a broad range of high frequencies. The difference between our case and these models is that we have an explicit breakdown of translational invariance due to the pinning potential. In the range of parameters where a pinned vortex glass exists, the pinning potential is expected to completely destroy the low-energy "phonon" part of the spectrum present in glasses~\cite{Boson1,Boson2}.
Indeed, this is what we observed numerically, see Extended Data Fig.~\ref{Fig_Hessian}: the density of states $\nu(\lambda)$ demonstrates a broad maximum similar to the "boson peak", between $\lambda_{\text{min}}=0$ and $\lambda_{\text{max}} \approx 2g^2$. The small number of negative eigenvalues seen in Extended Data Fig.~\ref{Fig_Hessian} results from a spurious technical reason, as explained in Supplementary Information IV; the shape of $\nu(\lambda >0)$ is insensitive to this spurious phenomenon.

\subsubsection{2D collective pinning regime}
In the case of a weak pinning potential, the vortex lattice is destroyed only at large spatial scales~\cite{Larkin1970,Larkin1979}, while short-range crystalline order is preserved locally. A theoretical description of the two-dimensional vortex lattice in this regime was developed in Ref.~\cite{FeigGeshLarkin1990} (see Sec.~5 therein). The characteristic size of a Larkin domain, i.e., the region of the vortex lattice that remains coherently pinned, can be estimated as~\cite{FeigGeshLarkin1990}
\begin{equation}
    R_p \approx \frac{g^2}{2W}\,\frac{r_c^2}{a_0},
    \label{Rc1}
\end{equation}
where $a_0$ is the intervortex spacing. In the collective pinning regime, the critical current density is then given by
\[
    j_c(B) \approx j_{c1}\frac{a_0}{R_p}
    \approx j_{c1}\frac{2W}{g^2}\frac{a_0^2}{r_c^2},
    \label{jc-collective}
\]
where $j_{c1}$ is the critical current density for independently pinned vortices. 
The curvature $\tilde{k}(B)$ of the effective pinning potential is related to $j_c(B)$ through
\[
\tilde{k}(B)\, r_c \approx \frac{\pi\hbar}{e}\, j_c(B)\, d,
\]
since $r_c$ sets the characteristic spatial scale of the pinning potential. As a result,
\begin{equation}
\tilde{k}(B) \approx \frac{\hbar}{e}\, j_{c1}\, d\, \frac{2\pi W}{g^2}\frac{a_0^2}{r_c^3}
\propto \frac{1}{B},
\label{k-collective}
\end{equation}
within the two-dimensional collective pinning scenario.

\subsection{Low magnetic fields: individual vortex pinning and estimate for $W$}
\label{LowB}

A random potential with variance $W$ and correlation length $r_c$ produces single-particle pinning minima with characteristic curvature of order $W/r_c^2$. More precisely, the vortex susceptibility is $\chi_v = g^2\alpha_v n r_c^2/W$, where $\alpha_v \approx 0.4$ according to our numerical data at small $n/n_c$ (see Supplementary Information).

At the crossover field $B^* \approx 2 \times 10^{-3} B_c$, the total suppression of $\theta(B)$ is $2\pi \chi_v \equiv 1 - \theta(B^*) = 0.2$ for samples DPRes2a and DPRes9a. Comparison with the above expression for $\chi_v$ yields the estimate
\begin{equation}
\frac{W}{g^2} = 5 \alpha_v \frac{B^*}{B_c} \frac{r_c^2}{\xi_0^2}
\approx 1.5 \times 10^{-2},
\label{W}
\end{equation}
where we use $r_c \approx 2\xi_0$, corresponding to $r_c/a \approx 0.8$. For the most disordered sample DPRes11a, a similar analysis gives $W/g^2 \approx 3.5 \times 10^{-2}$.

To cross-check these estimates against experiment, we evaluate the critical current density in the low-$B$ regime by equating the Lorentz force to the pinning force,
\[
j_c \frac{\pi\hbar}{e} d \sim \frac{W}{r_c}.
\]
Using the parameters for sample DPRes8a (which is similar to DPRes2a) from Extended Data Table~\ref{tab:SampleParameters}, we obtain
\[
j_c \sim 4 \times 10^{4} \frac{W}{g^2}\,\mathrm{A/cm^2}.
\]
This estimate exceeds the $B \to 0$ extrapolation of the data for DPRes8a in Extended Data Fig.~\ref{Fig5} by only a factor of approximately 1.5.

It may appear surprising that a value $W/g^2 \approx 1.5 \times 10^{-2}$ already corresponds to strong pinning. The following estimates, which build on the qualitative discussion of Ref.~\cite{BraggGlass}, clarify this point.

For the expression in Eq.~(\ref{Rc1}) to be meaningful, the ratio $R_p/a_0$ must be large compared to unity. At low magnetic fields, vortices are pinned individually and their positions follow the random potential landscape; typical displacements of neighboring vortices relative to their nominal spacing $a_B$ are of order $r_c$. Upon increasing $B$, two scenarios are possible. 

If the ratio $R_p/a_B$ exceeds unity while $r_c/a_B$ remains smaller than the Lindemann criterion $c_L \approx 0.15$, the system evolves from single-vortex pinning into a collectively pinned vortex lattice, as described in Refs.~\cite{FeigGeshLarkin1990,GlD1995}. By contrast, if $r_c/a_B$ reaches $c_L$ while $R_p/a_B$ is still small, crystalline order is lost before collective pinning can develop, and the system instead enters the pinned vortex glass (PVG) regime~\cite{BraggGlass}.

The boundary between these two regimes occurs at $W/g^2 = c_L^2/8 \approx 2.8 \times 10^{-3}$, which is substantially smaller than the disorder strength estimated above for a:InO films. We therefore expect a direct crossover from the single-vortex regime to the PVG regime. In contrast, MoGe films, characterized by much weaker pinning, fall within the collective pinning (Bragg glass) scenario.

\subsection{Estimates of strong and ultrastrong coupling in magnetic field }
\label{method:strong_coupling}
Due to the increase in kinetic inductance $L_K(B)$ when approaching the superconductor-insulator transition, the resonator's characteristic impedance $Z(B) = \sqrt{L_K(B)/C}$ is strongly enhanced with magnetic field (Extended data Fig.~\ref{Ext_Fig5}a). For our $\lambda/2$ resonator geometry, $Z$ can be computed from $Z = 2 f(B) L_K^{\square}(B) (l/w)$, where $l$ and $w$ are the resonator length and width (see Extended Data Table~\ref{tab:SampleParameters}) and $f(B)$ is the resonance frequency. Extended Data Fig.~\ref{Ext_Fig5}b shows this impedance enhancement for three resonators of different disorders and geometries. Near the transition, $Z$ largely exceeds the resistance quantum for Cooper pairs $R_Q = h / (2e)^2 \approx 6.45~\mathrm{k \Omega}$ (shown by the horizontal dashed line in Extended Data Fig.~\ref{Ext_Fig5}b). At a given disorder level, the impedance can be controlled by varying the resonator width, as shown in Extended data Fig.~\ref{Ext_Fig5}. 
Regarding dissipation, the resonator internal quality factors remain finite (albeit small) up to multitesla out-of-plane magnetic fields. Extended Data Fig.~\ref{Ext_Fig4}b shows how $Q_i(B)$ plateaus at $Q_i \sim 100$ up to about $1$ Tesla, and slowly decays at higher fields upon approaching the SIT.

Together, field-tunable, ultra-high impedance and preserved mode visibility at large out-of-plane fields make disordered superconducting resonators near the SIT an interesting platform for quantum sensing~\cite{Clerk2020}.
In particular, high impedances give rise to strong zero-point fluctuations of the voltage along the resonator, given by $V_{\text{zpf}}(B) = \omega_r(B) \sqrt{\hbar Z(B)/2}$, thereby facilitating the coupling to quantum systems via the relation $ \hbar g = \eta e V_{\text{zpf}}$ where $g$ is the coupling strength and $\eta \leq 1$ is a dimensionless factor characterizing the coupling efficiency between the dipole and the resonator's electric field, which depends on the geometry and the nature of the coupled system.

Our resonators make the ultra-strong coupling (USC) regime in magnetic field experimentally accessible, defined by $g/\omega_r \gtrsim 0.1$~\cite{FornDiaz2019,FriskKockum2019}. Since $g/\omega_r \approx 0.88\,\eta\sqrt{Z/R_Q}$ (the prefactor $\sqrt{\pi}/2 \approx 0.88$ stems from the $\lambda/2$ geometry), Extended Data Fig.~\ref{Ext_Fig5}b indicates that the USC regime at 8-9~T could be achieved for $\eta \gtrsim 0.1$ in sample DPres2. Such coupling strengths are readily accessible in systems with large electric dipole moments or strong spin--orbit interaction, including hole spin qubits or nanomechanical resonators. Strong coupling, defined as $g/ \kappa = (g/\omega_r) Q_L \approx 0.88 \, \eta \, Q_L\sqrt{Z/R_Q} > 1$ where $Q_L$ is the loaded quality factor, is also enabled by our resonators, especially at $B\sim 1$ T where the dissipation remains low. Here, $\eta$ should satisfy the condition $\eta > 0.005 - 0.01$, which is readily achievable with, for example, a capacitively coupled double quantum dot.
In principle, the impedance could be further enhanced by an order of magnitude by reducing the resonator width which would further relax the conditions on coupling efficiency.


\bigskip

\section*{Acknowledgments}
We thank Vadim Geshkenbein, Anton Khvalyuk and Igor Poboiko for valuable discussions. 
Samples were prepared at the Nanofab facility of the N\'eel Institute. D.P. and B.S acknowledge funding from the ANR agency under 11 the 'France 2030' plan, with Reference No. ANR-22-PETQ-0003. B.S. has received funding from the European Union's Horizon 2020 research and innovation program under the ERC grant SUPERGRAPH No. 866365. 
M.F. acknowledges a collaboration with ICTP (R. Fazio) under European Union (ERC, RAVE, 101053159).

\section*{Author contributions} D.P., T.C., N.K., and K.R. performed the measurements. D.P., T.C., N.K., N.B., and X.P. designed and fabricated the samples.  D.P., T.C., N.K., K.P. and B.S. performed the data analysis. F.B. and F.G. worked on the measurement setup. M.F. and V.K. developed the theory and contributed to the data analysis. V.K. developed the code and performed the numerical simulations. B.S. conceived the project and supervised it. B.S., T.C., V.K. and M.F. wrote the paper with inputs from all co-authors.

\section*{Competing Interests} The authors declare that they have no competing interests.

\bibliography{biblio}

\newpage

\setcounter{figure}{0}
\renewcommand{\figurename}{Extended Data Fig.}
\renewcommand{\tablename}{Extended Data Table}

\begin{table*}[t]
	\centering
		\begin{tabular}{|c|c|c|c|c|c|c|c|c|c|c|}
		 \hline
			Sample & $w~(\upmu \mathrm{m})$ & $l~(\mathrm{mm})$ & $d~(\mathrm{nm})$ 
            & $\mathcal{W}~(\upmu \mathrm{m})$ & $L~(\upmu \mathrm{m})$ & $T_{\rm c}~(\mathrm{K})$ & $B_{\rm c}~(\mathrm{T})$ & $R_{\square}(\mathrm{k\Omega/\square})$ & $L_{\rm K}~(\mathrm{nH/\square})$ & $B_{\rm c}^{\Theta}~(\mathrm{T})$  \\
            			\hline
			DPRes2a & 6 &  3.76 & 57 & - & - & - & - & - & 1.34 & 9.7  \\
			\hline
            DPRes2b & 4  &  2.7 & 57 & - & - & - & - & - & 1.34& 9.3  \\
			\hline
            DPRes3a & 10 & 4.7 & 53 & - & - & - & - & - & 1.77 & -  \\
            \hline
            DPRes3b & 1 & 1.4 & 53 & - & - & - & - & - & 1.78& -  \\
            \hline
            DPRes8a & 8 & 4.17 & 70 & 6.0 & 24.0 & 2.7 & 10.4  & - & 1.0 & 11.3   \\
			\hline
            DPRes8b & 2 & 1.9 & 70 & 6.0 & 24.0 & 2.7 & 10.4  & - & 1.0 & 11.2   \\
            \hline
            DPRes9a & 5 & 0.77 & 44 & 6.0 & 24.0 & 2.2 & 7.0 & 5.5 & 3.1 & 5.6   \\
			\hline
            DPRes9c & 10 & 0.59 & 44 & 6.0 & 24.0 & 2.2 & 7.0 & 5.5 & 2.97 & 5.7   \\
            \hline
			DPRes11a & 5 & 0.77 & 47 & 10.0 & 30.0 & 0.94 & 0.4 & 10.65 &11.1 & 0.4   \\
			\hline
			DPRes11c & 10 & 0.59 & 47 & 10.0 & 30.0 & 0.94 & 0.4 & 10.65 & 10.1& 0.4   \\
			\hline
            DPRes13c & 10 & 0.59 & 48 & 10.0 & 30.0 & 1.1 & 0.95 & 7.4 &8.38 & 1.0   \\
			\hline
            DPRes13a & 5 & - & 48 & - & 1.1 & 1.3 & 1.3 & 8  & 8.56 & - \\
			\hline
            NK004 & 1 & 6.9 & 10 & 1 & 29 & - & - & -  & 0.07 & 8.8 \\
			\hline
		\end{tabular}
    \caption{\textbf{Sample parameters.}
    $w$, $l$ and $d$ denote the resonator width, length and thickness, respectively. $\mathcal{W}$ and $L$ denote the width and length of the d.c. transport mesa strip. The superconducting critical temperature $T_{\mathrm{c}}$, the critical magnetic field $B_{\mathrm{c}}$ and the sheet resistance $R_{\square}$ are extracted from transport measurements. $R_{\square}$ is defined as the maximum sheet resistance measured just above $T_{\mathrm{c}}$. $B_{\mathrm{c}}^{\Theta}$ and $L_{\mathrm{K}}$ are extracted from microwave resonance measurements. $L_{\mathrm{K}}$ is obtained by comparing the zero-field resonance frequency with electromagnetic simulations.}     
	\label{tab:SampleParameters}
\end{table*}

\begin{figure}[ht!]
\includegraphics[width=0.5\textwidth]{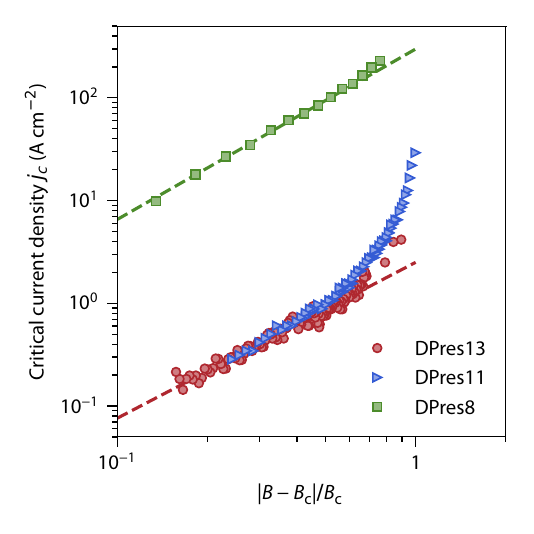}
\centering
\caption{\textbf{Critical current densities near $B_c$.}
Critical current density as a function of the reduced distance to the critical field, $|B-B_c|/B_c$, for three different samples. The critical fields are $B_c = 0.95~\mathrm{T}$, $0.35~\mathrm{T}$ and $10.4~\mathrm{T}$ for samples DPRes13, DPRes11 and DPRes8, respectively. The dashed red and green lines are fits to the power-law dependence yielding exponents 1.5 and 1.6, respectively.}
\label{Fig5}
\end{figure}

\begin{figure*}[ht!]
\includegraphics[width=0.9\textwidth]{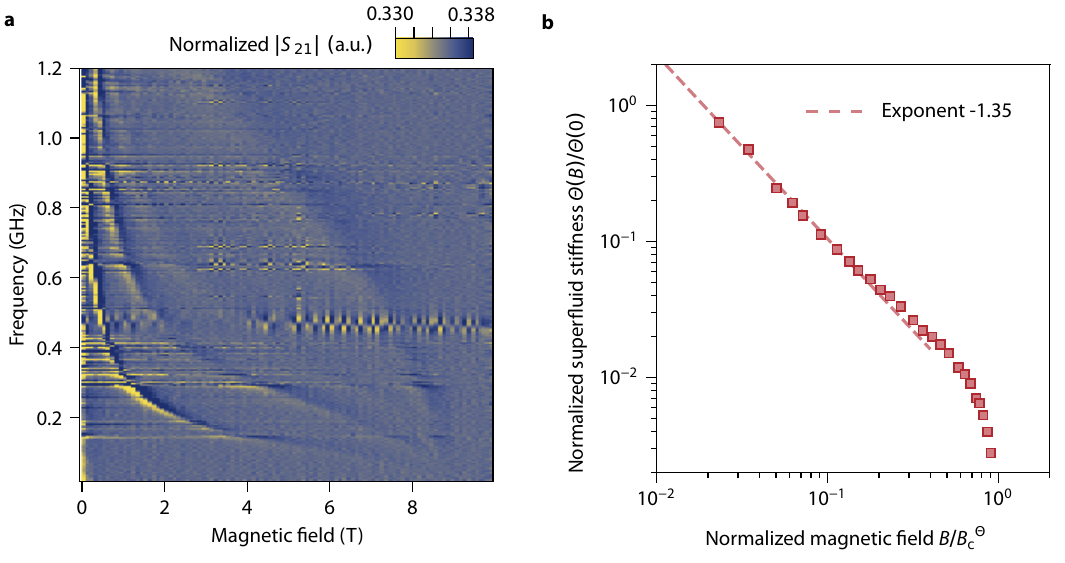}
\centering
\caption{\textbf{Superfluid density of low-disorder MoGe resonators.}
\textbf{a,} Microwave transmission $|S_{21}|$ as a function of magnetic field and frequency for the low-disorder MoGe sample NK004, comprising two resonators. Owing to significant measurement noise and the presence of numerous field-independent parasitic modes, visible as horizontal lines, the data were interpolated and differentiated to remove linear backgrounds, then normalized and smoothed using a median filter. 
\textbf{b,} Normalized superfluid stiffness $\Theta(B)/\Theta(0)$ as a function of normalized magnetic field, $B/B_c^{\Theta}$, with $B_c^{\Theta} = 8.8~\mathrm{T}$. The dashed line is a fit to $\Theta(B)/\Theta(0) \propto (B/B_c^{\Theta})^{-1.35}$.}
\label{Ext_Fig2}
\end{figure*}

\begin{figure*}[ht!]
\includegraphics[width=\textwidth]{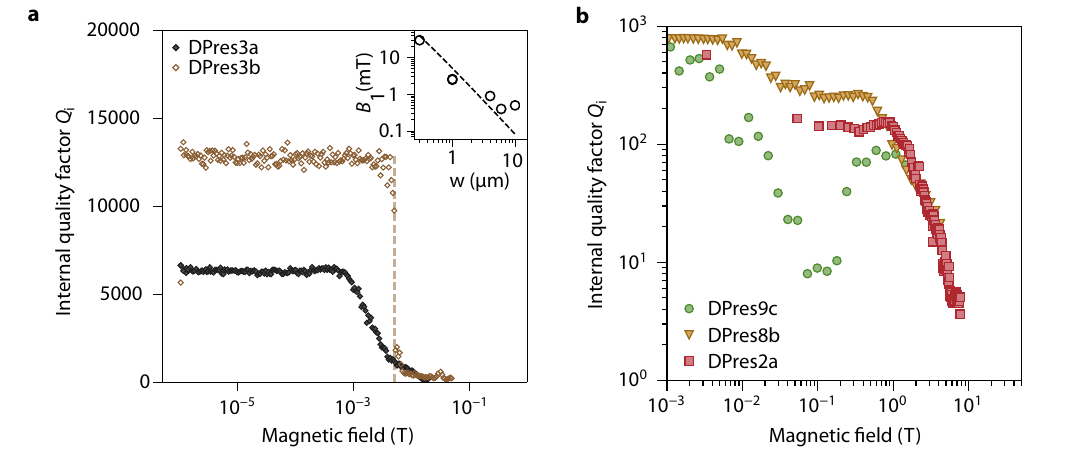}
\centering
\caption{\textbf{Microwave dissipation in out-of-plane magnetic field.}
\textbf{a,} Internal quality factor $Q_{\mathrm{i}}$ as a function of applied perpendicular magnetic field for two resonators of different widths on sample DPRes3: DPRes3a is $10~\upmu\mathrm{m}$ wide, whereas DPRes3b is $1~\upmu\mathrm{m}$ wide. The sudden drop in quality factor results from vortex entry, which occurs at the threshold field $B_1$, as discussed in Methods~\ref{sec:vortex_entry}. The vertical dashed line indicates the calculated value of $B_1$ for a strip with $w = 1~\upmu\mathrm{m}$ and $\xi = 5~\mathrm{nm}$. Inset: threshold magnetic field for vortex entry, defined from the drop in quality factor, as a function of resonator width. The dashed line is the theoretical value of $B_1$ from Eq.~(\ref{equ1}), calculated with $\xi = 5~\mathrm{nm}$.
\textbf{b,} $Q_{\mathrm{i}}$ at higher magnetic fields for samples DPRes9, DPRes8 and DPRes2, which have different disorder levels and widths. Upon approaching $B_{c2}$, the three curves converge, suggesting a disorder- and geometry-independent dissipation mechanism.}
\label{Ext_Fig4}
\end{figure*}

\begin{figure*}[ht!]
\includegraphics[width=\textwidth]{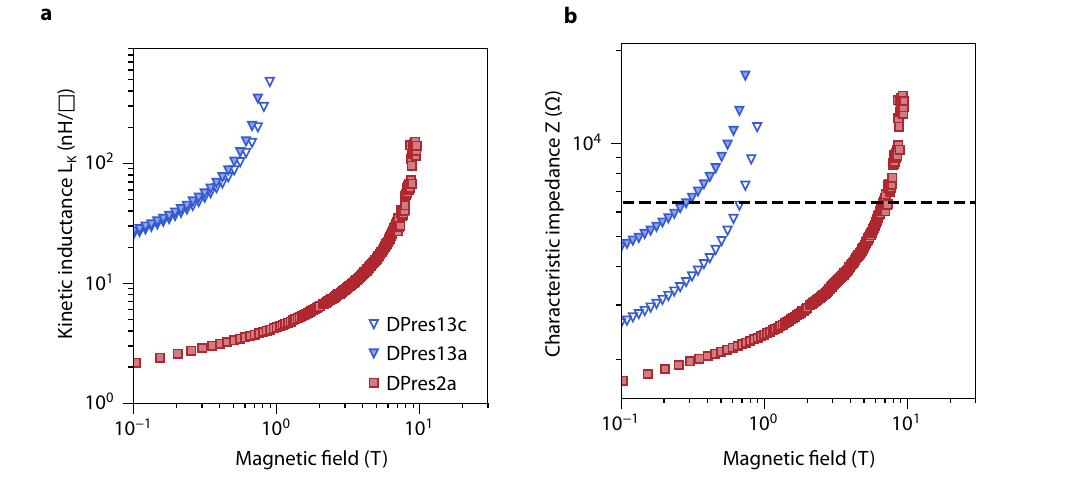}
\centering
\caption{\textbf{Diverging kinetic inductance and impedance near the SIT.}
\textbf{a,} Sheet kinetic inductance as a function of magnetic field for three resonators near the superconductor--insulator transition. Two resonators of different widths are shown for the strongly disordered sample DPRes13: DPRes13c is $10~\upmu\mathrm{m}$ wide and DPRes13a is $5~\upmu\mathrm{m}$ wide. One resonator is shown for the moderately disordered sample DPRes2.
\textbf{b,} Characteristic impedance, $Z = \sqrt{L_K/C}$, of the same resonators as a function of magnetic field. The horizontal dashed line indicates the Cooper-pair resistance quantum, $R_Q = h/(2e)^2 \simeq 6.45~\mathrm{k}\Omega$.}
\label{Ext_Fig5}
\end{figure*}

\begin{figure}[!ht]
\includegraphics[width=0.9\columnwidth]{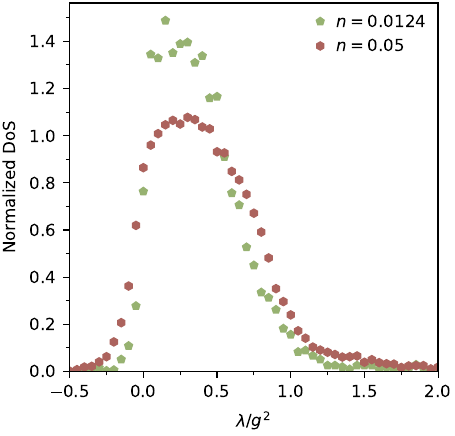}
\centering
\caption{Computed normalized density of states $\nu(\lambda)$ versus Hessian eigenvalues (in units of $g^2$), for $W=0.5 g^2$ and two vortex densities $n = 0.0124$ and $n=0.05$, averaged over 100 and 50 realizations of disorder, respectively.}
\label{Fig_Hessian}
\end{figure}

\bigskip
\clearpage
\bigskip

\section*{Supplementary Information}

\setcounter{figure}{0}
\renewcommand{\figurename}{Fig.}
\renewcommand{\thefigure}{S\arabic{figure}}

\section{Numerical method}
\label{numerical}

\subsection{Numerical study of vortex glass.}

In order to simulate the vortex glass, we introduce a square lattice with lattice constant $a$, on which vortices can reside. Typically, we consider systems of size $L \times L$ with $L=50$ and $100$. The interaction between vortices is described by

\begin{equation}
V(\mathbf{r},\mathbf{r}^{'})=
\begin{cases}
- v\ln{(2|\mathbf{r}-\mathbf{r}^{'}|/L)} & \text{if } |\mathbf{r}-\mathbf{r}^{'}|<L/2\\
0 & \text{if } |\mathbf{r}-\mathbf{r}^{'}|>L/2
\end{cases}
\label{Vnum1}
\end{equation}

In all simulations, we use $v=1$, which corresponds to the parameter $g^2 =1$ (see \textit{Methods}, Sec.~F).

The on-site energies $\epsilon(r_i)$ are random variables with a Gaussian distribution function:

\begin{equation}
P(\epsilon)=(2\pi W)^{-1/2}\exp{(-\epsilon^2/2W^2)}.
\end{equation}

Spatial correlations are determined by the correlation function:

\begin{equation}
\overline{\epsilon(r)\epsilon(r^{'})}=W^2\exp{(-|\mathbf{r}-\mathbf{r}^{'}|^2/r_c^2)},
\end{equation}
where $W$ describes the width of the energy distribution; we choose it in the range $0.05 < W \leq 0.8$. The parameter $r_c$ is the correlation radius characterizing the spatial correlations of the energy distribution. Typically, $r_c$ varies within the interval $0.5 < r_c < 2$.

The spatially correlated disorder is generated using the NAG library routine \texttt{g05zrf}. For calculating the interaction between vortices and for vortex motion, we assume periodic boundary conditions.

In order to find a local minimum, we apply a slightly modified version of the method used in the theory of Coulomb glasses \cite{Glatz2008}. As in Ref.~\cite{Glatz2008}, our minimization procedure consists of three steps.

In the first step, exchanges between neighboring states are performed. Pairs of occupied sites are selected randomly, and pairwise exchanges that reduce the total energy are accepted. Once no further nearest-neighbor exchanges can lower the energy, the second step of the minimization is performed.

In the second stage of the algorithm, we select the occupied site with the largest addition energy $\Delta\epsilon_{\mathrm{max}}$ and move the vortex to the empty site with the minimal addition energy $\Delta\epsilon_{\mathrm{min}}$. This stage is repeated as long as $\Delta\epsilon_{\mathrm{max}} > \Delta\epsilon_{\mathrm{min}}$.

In the third step, the occupied site with the maximal addition energy is exchanged with the empty site that provides the maximal total energy gain. Only vortex moves that reduce the total energy are accepted. After the third step is completed, the procedure starts again and is repeated until no further pair exchange is possible. The resulting configurations are defined as local minima.

To calculate the susceptibility $\chi$, we apply to the system (after finding a local minimum as described above) an inhomogeneous force consistent with periodic boundary conditions. This is implemented by adding the following extra energy term to each site:
\begin{equation}
\tilde{\epsilon}(r_i)=\epsilon(r_i)-f\cos{(2\pi x_i/L)},
\end{equation}
where $f$ is the strength of the external force. In our simulations, we use values in the range $0.5 \leq f \leq 7$. These forces are sufficiently strong to produce a measurable change in the total energy, while still being small enough to remain within the linear-response regime.

\begin{figure}[!ht]
\centering
\includegraphics[width=1\columnwidth]{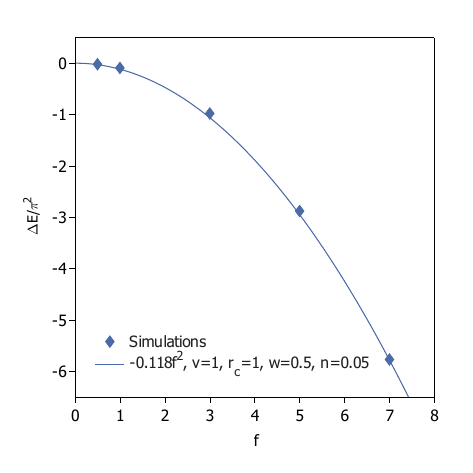}
\caption{$\Delta E=E(f)-E(0)$ as a function of the force $f$. The results are well fitted by a parabola.}
\label{theoryFig0}
\end{figure}

As a result, the total energy after applying the force takes the form

\begin{equation}
E(f)=E(0)-\chi L^2\overline{F(r)^2}/2
     =E(0)-\pi^2 f^2 \chi,
\end{equation}
where the force is given by
\begin{equation}
F(r)=-\frac{2\pi}{L}\sin(2\pi x_i/L).
\end{equation}
Therefore,
\begin{equation}
\chi=-\frac{E(f)-E(0)}{\pi^2f^2}.
\end{equation}
In Fig.~\ref{theoryFig0} we plot the change in energy $\Delta E=E(f)-E(0)$ as a function of the force constant $f$. The results show that the energy change scales as $f^2$ up to $f=7$. This indicates that, for the case $W=0.5$, the system remains in the linear-response regime up to this value of the applied force.

Once the susceptibility $\chi_a$ is calculated for a given disorder configuration $\{a\}$, we can determine the inverse dielectric constant using the relation (see \textit{Methods}, Sec.~F)

\begin{equation}
\varepsilon^{-1} \equiv \left\langle \frac1{\varepsilon_a} \right\rangle
=
\left\langle \frac{1}{1+ 2\pi\chi_a} \right\rangle,
\label{eps-a}
\end{equation}
where $\langle \cdots \rangle$ denotes averaging over disorder realizations.
As we show below, both $1/\varepsilon$ and $\chi = \langle \chi_a \rangle$ depend linearly on $\ln(n)$. Therefore, the simple relation
$1/\varepsilon = 1/(1 + 2\pi \chi)$ does not hold.
Typically, averaging over 100 disorder realizations is sufficient to calculate $\chi$ and $\varepsilon$ with a statistical error below $5\%$.

An additional free parameter employed in the computation of the susceptibility is the maximal hopping distance allowed for a particle while calculating the response to the external force $f$.
Since we are ultimately interested in the high-frequency (GHz) response, it is reasonable to consider relatively short hops, as longer hops would require excessively long times in our glassy system. We therefore consider either hopping within two coordination shells, corresponding to hops of length $a$ and $\sqrt{2}a$, or within three shells, additionally including hops of length $2a$.
This point will be discussed at the end of the next section, where the dependence on the number of allowed hopping shells will be demonstrated.

\subsection{Determination of the upper critical field within the numerical model}

To establish a detailed connection between the experimental results and the numerical simulations, we need to identify the quantity $B_c^{\mathrm{num}}$ which, within our numerical algorithm, plays the same role as the upper critical field $B_{c2}$ in a real superconductor. Although this relation is qualitatively evident,
\[
B_c^{\mathrm{num}} \sim \Phi_0/a^2,
\]
we would like to determine the numerical prefactor with reasonable accuracy. 
Below we describe the procedure used to achieve this goal.
In this part of the work, we do not require a random potential, and therefore we set $W=0$.

To implement this program, we introduce a long-range cutoff $\lambda_1$ for the logarithmic interaction in Eq.~(\ref{Vnum1}):

\begin{equation}
V_2(r) = K_0(r/\lambda_1),
\label{Vnum2}
\end{equation}
where we choose $\lambda_1 = L$. Such an interaction between point particles with unit ``charges'' follows from the energy functional
\begin{equation}
E\left[\psi(\mathbf{r})\right] =
2\pi\lambda_1^2 \int d^2r
\left[
\frac12\left(\psi^2 + \lambda_1^2 (\nabla\psi)^2 \right)
- \sum_j\psi\delta(\mathbf{r}-\mathbf{r}_j)
\right].
\label{Epsi}
\end{equation}
Here, the points $\mathbf{r}_j$ correspond to vortex positions, and the summation runs over all vortices, which are assumed to form a conventional triangular lattice (slightly distorted due to the underlying square grid with lattice spacing $a$). The elementary side length of the triangular lattice is
\[
a_\triangle = \left(\frac{2}{n\sqrt{3}}\right)^{1/2},
\]
and throughout this analysis we assume $a_\triangle \gg a$.

Transforming to Fourier space, we obtain, in terms of the reciprocal lattice vectors $\mathbf{K}$,
\begin{equation}
E\left[\psi_{\mathbf{K}}\right] = - \pi\lambda_1^2\sum_{\mathbf{K}} \frac{n^2}{1 + \lambda_1^2 \mathbf{K}^2}
=
- \pi\lambda_1^2 n^2
- \frac{n}{2}\ln\frac{\beta a_\triangle}{a},
\label{Epsi2}
\end{equation}
where $n$ is the areal density of vortices and $\beta \sim 1$ is a numerical coefficient to be determined later.
The second term on the right-hand side of Eq.~(\ref{Epsi2}) arises from estimating the sum over $\mathbf{K} \neq 0$ and transforming it into a logarithmic integral between $1/a_\triangle$ and $1/a$ under the condition $\lambda_1 \gg a_\triangle$.

We now define the quantity
\[
h \equiv - \frac{1}{2\pi\lambda_1^2}\frac{dE}{dn},
\]
and evaluate it using Eq.~(\ref{Epsi2}):

\begin{equation}
h =
n +
\frac{1}{8\pi \lambda_1^2}
\ln\frac{2\beta^2}{\sqrt{3} e a^2 n}
\equiv
n +
\frac{1}{8\pi \lambda_1^2}
\ln\frac{n_c}{n},
\label{h1}
\end{equation}
where $e = 2.718\ldots$ and
\[
n_c a^2 \equiv \frac{2\beta^2}{e\sqrt{3}}.
\]
The quantities $n$ and $h$ in Eq.~(\ref{h1}) play roles analogous to the magnetic induction $B$ and magnetic field $H$ in the standard description of the mixed state of type-II superconductors in the range
\[
H_{c1} \ll B \ll H_{c2},
\]
which corresponds to the inequalities
\[
1/\lambda_1^2 \ll n \ll 1/a^2.
\]
The second term on the right-hand side of Eq.~(\ref{h1}) represents the diamagnetic magnetization $M$ of the mixed state, which vanishes at the upper critical field $B=H_{c2}$, corresponding to $n=n_c$.

We numerically evaluate the energy $E[n]$ of vortices on a large lattice of size $L=128$, determine the coefficient $\beta=1.5$, and ultimately obtain
\[
n_c = a^{-2}\cdot \frac{2\beta^2}{e\sqrt{3}}
= \frac{0.96}{a^2}.
\]

Since
\[
n_c = \frac{B_{c2}}{\Phi_0}
\equiv \frac{1}{2\pi\xi^2},
\]
this allows us to establish the relation
\[
a \approx 2.5\xi.
\]

\section{Logarithmic field dependence: results of numerical simulations}
\label{logarithm}

\begin{figure}[!ht]
\includegraphics[width=1\columnwidth]{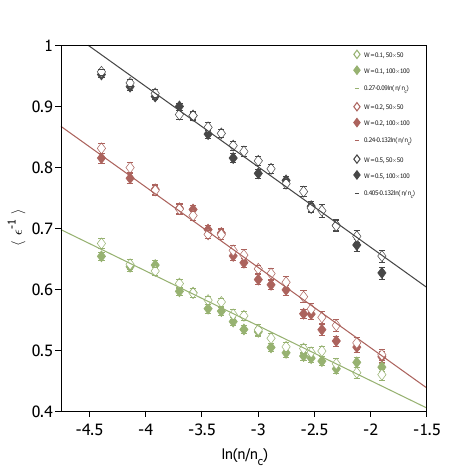}
\centering
\caption{Average inverse permittivity $\langle\varepsilon^{-1}\rangle$ calculated for $W=0.1$, $0.2$, and $0.5$, with correlation length $r_c=1$. Open symbols correspond to system size $L=50$, while solid symbols correspond to $L=100$. The results are fitted by the formula $\langle\varepsilon^{-1}\rangle=-A\ln(n/n_c)+B$, with $A=0.09$ and $B=0.27$ for $W=0.1$, $A=0.132$ and $B=0.24$ for $W=0.2$, and $A=0.132$ and $B=0.405$ for $W=0.5$.}
\label{theoryFig1}
\end{figure}

Typical simulation results for the average inverse dielectric permittivity $\langle\varepsilon^{-1}\rangle$ are presented in Fig.~\ref{theoryFig1} for different disorder strengths, $W=0.1$, $0.2$, and $0.5$. The dependence is linear in $\ln(n/n_c)$ over a broad range of vortex concentrations, except at the lowest densities. Here we use $n_c=1/a^2$. Solid symbols correspond to calculations performed on systems of size $L=100$, while open symbols correspond to $L=50$.

The simulations demonstrate that the results for $L=50$ and $L=100$ are very close to each other, indicating that the inverse permittivity is essentially independent of system size.

\begin{figure}[!ht]
\includegraphics[width=1\columnwidth]{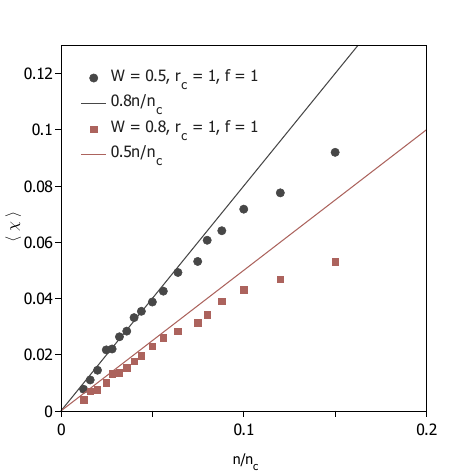}
\centering
\caption{Susceptibility calculated for $r_c=1$ and different values of $W$ as a function of vortex density ($W=0.5$ shown by black points and $W=0.8$ by red points). In the low-density regime, the results are well approximated by linear dependences with slopes $0.8$ for $W=0.5$ and $0.5$ for $W=0.8$, respectively.}
\label{theoryFig2}
\end{figure}

Although the logarithmic dependence on $n$ clearly dominates, it is also useful to examine the low-density regime in more detail. Figure~\ref{theoryFig2} shows the susceptibility as a function of vortex density $n$ for $r_c=1$ and $f=1$, for different values of $W$, averaged over 100 disorder realizations.
The graph clearly demonstrates a linear dependence of the averaged susceptibility $\langle\chi\rangle$ on $n$ in the low-density regime. This behavior is consistent with the relation

\begin{equation}
\chi_v = \alpha_v\frac{nr_c^2}{W},
\label{chi-f}
\end{equation}
and allows us to estimate the coefficient $\alpha_v$. For example, for $W=0.5$, the slope in Fig.~\ref{theoryFig2} is approximately $0.8$, yielding $\alpha_v=0.4$. The same value, $\alpha_v=0.4$, is obtained from the data for $W=0.8$.

\begin{figure}[!ht]
\includegraphics[width=1\columnwidth]{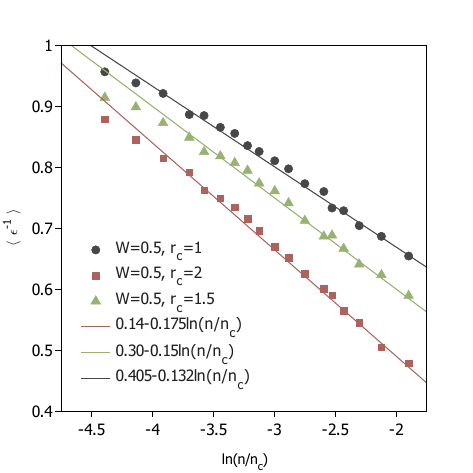}
\centering
\caption{Inverse dielectric function $\langle\varepsilon^{-1}\rangle$ for $W=0.5$ and different values of $r_c$ as a function of $\ln(n/n_c)$.}
\label{theoryFig3}
\end{figure}

Figure~\ref{theoryFig3} demonstrates the dependence of the simulation results on the correlation length $r_c$. Here we plot the inverse dielectric function $\langle\varepsilon^{-1}\rangle$ calculated for $W=0.5$ and different values of $r_c$ as a function of $\ln(n/n_c)$, averaged over 100 disorder realizations.

The inverse dielectric function exhibits a linear dependence on $\ln(n/n_c)$. The slope
\[
\mathcal{S} = - \frac{d\langle\varepsilon^{-1}\rangle}{d \ln(n/n_c)}
\]
increases with increasing $r_c$.

\begin{figure}[!ht]
\includegraphics[width=1\columnwidth]{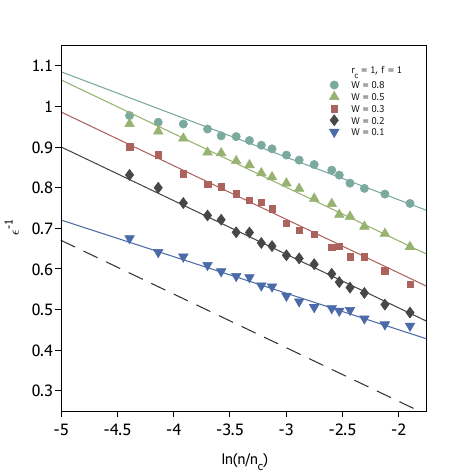}
\centering
\caption{Inverse dielectric function calculated for $r_c=1$ and $f=1$: $W=0.2$ (black points), $W=0.3$ (red points), $W=0.5$ (green points), $W=0.1$ (blue points), and $W=0.8$ (cyan points). The results are fitted by the formula $\langle\varepsilon^{-1}\rangle=- \mathcal{S}\ln(n/n_c)+B$ with $\mathcal{S} =0.132$ for $W=0.2$, $0.3$, and $0.5$; $\mathcal{S}(W=0.8)=0.105$; and $\mathcal{S}(W=0.1)=0.09$. The corresponding offset coefficients are $B=0.24$, $0.326$, $0.405$, $0.56$, and $0.2$ for $W=0.2$, $0.3$, $0.5$, $0.8$, and $0.1$, respectively. The dashed line indicates the position of the experimental data shown in Fig.~2 of the Main Text.}
\label{theoryFig4}
\end{figure}

On the other hand, the dependence of the slope $\mathcal{S}$ on $W$ is relatively weak over a broad range of disorder strengths, as shown in Fig.~\ref{theoryFig4}.

In Fig.~\ref{theoryFig4}, we plot the averaged inverse dielectric function as a function of $\ln(n/n_c)$ for $r_c=1$, $f=1$, and different disorder strengths $W$. The figure demonstrates that the slope $\mathcal{S}$ is nearly independent of $W$ over a relatively broad disorder range.

The sharp decrease of the slope for $W=0.1$ is attributed to vortex depinning, since the force magnitude $f=1$ becomes too strong for such weak disorder.

\begin{figure}[!ht]
\includegraphics[width=1\columnwidth]{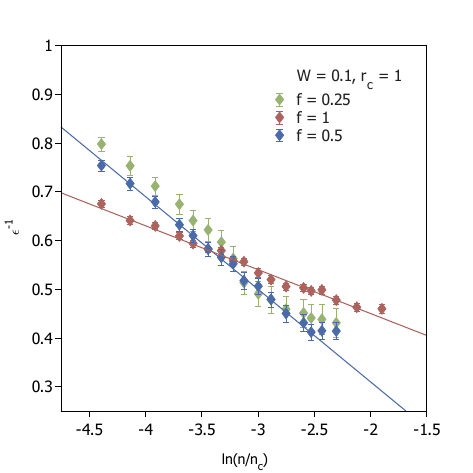}
\centering
\caption{Inverse dielectric function $\langle\varepsilon^{-1}\rangle$ for $W=0.1$ and different values of $f$ as a function of $\ln(n/n_c)$. The results are fitted by the formula $\langle\varepsilon^{-1}\rangle=- \mathcal{S}\ln(n/n_c)+B$ with $\mathcal{S} =0.09$ for $f=1$ and $\mathcal{S} =0.19$ for $f=0.5$ and $f=0.25$. The corresponding offset coefficients are $B=0.27$ for $f=1$ and $B=-0.07$ for $f=0.25$ and $0.5$.}
\label{theoryFig6}
\end{figure}

To support the above interpretation, we performed simulations for several different values of the force constant $f$, as shown in Fig.~\ref{theoryFig6}. The slope $\mathcal{S}$ decreases significantly when $f$ increases from $0.5$ to $1$. In contrast, the slope remains nearly unchanged between $f=0.25$ and $f=0.5$.

\begin{figure}[!ht]
\includegraphics[width=1\columnwidth]{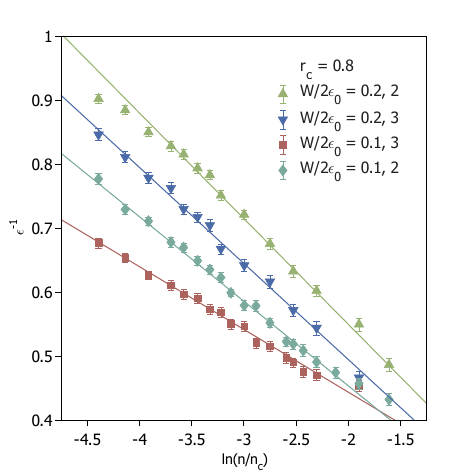}
\centering
\caption{Inverse dielectric function $\langle\varepsilon^{-1}\rangle$ for $W=0.1$ and $W=0.2$, calculated with maximal hopping distances extending to either the second (2) or third (3) coordination shell, plotted as a function of $\ln(n/n_c)$. The results are fitted by the formula $\langle\varepsilon^{-1}\rangle=- \mathcal{S}\ln(n/n_c)+B$. For $W=0.2$, we obtain $\mathcal{S}=0.165$ and $B=0.22$ for hopping up to the second shell, and $\mathcal{S}=0.15$ and $B=0.195$ for hopping up to the third shell. For $W=0.1$, we obtain $\mathcal{S}=0.132$ and $B=0.019$ for hopping up to the second shell, and $\mathcal{S}=0.098$ and $B=0.248$ for hopping up to the third shell.}
\label{theoryFig7}
\end{figure}

Finally, we comment on the dependence of the averaged inverse dielectric constant on the maximal hopping length. In Fig.~\ref{theoryFig7}, we plot the averaged inverse dielectric constant as a function of $\ln(n/n_c)$ for different hopping ranges.
As expected, the inverse dielectric constant decreases when the maximal hopping distance increases from the second coordination shell to the third coordination shell. In addition, the slope $\mathcal{S}$ decreases slightly as the hopping range increases.

The experimental data shown in Fig.~2 of the Main Text (red squares and green circles) are well approximated by the dependence
\[
\langle\varepsilon^{-1}\rangle
=
-\mathcal{S}\ln(B/B_{c2})+B,
\]
with $\mathcal{S}=0.13$ and $B=0.01$.
Our numerical simulations indicate that the slope $\mathcal{S}$ depends primarily on the correlation length $r_c$, and we therefore choose $r_c$ to fit the experimental data. Our primary choice, used throughout the Main Text and Methods section, is
\[
r_c = 0.8a = 2\xi_0.
\]

\begin{figure}[!ht]
\includegraphics[width=1\columnwidth]{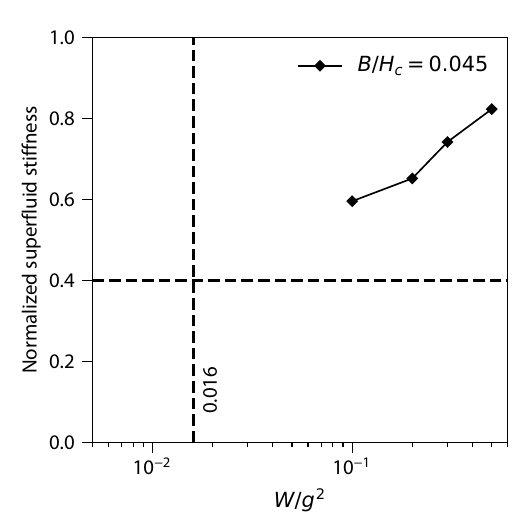}
\centering
\caption{Simulated normalized superfluid stiffness as a function of $W/g^2$ at the normalized magnetic field $B/H_c = 0.045$. The vertical and horizontal dashed lines correspond to $W/g^2 = 0.016$ and $\Theta/\Theta(0) = 0.4$, respectively.}
\label{FigS8}
\end{figure}

To further compare our simulations with the experimental data, we note that for the samples DPres2 and DPres9, the parameter $W/g^2$ is approximately $0.016$ and $0.003$, respectively (see Main Text). Since our simulations cannot reliably resolve such small values of $W/g^2$, we present in Fig.~\ref{FigS8} the normalized superfluid stiffness as a function of $W/g^2$ on a logarithmic $x$-axis scale, for the representative normalized magnetic field $B/H_c = 0.045$.
At this magnetic field the experimental value is approximately $0.4$, as shown in the inset of Fig.~2. Therefore, the intersection of the lines $W/g^2 = 0.016$ and $\Theta/\Theta(0) = 0.4$ 
indicates the expected prediction of our theory extrapolated to small enough $W$. 
It is evident from Fig.~\ref{FigS8} that the four simulation points converge toward this crossing point, supporting the quantitative agreement between our model and the experimental data.

\section{Numerical simulations with other types of interaction between vortices}
\label{screened}

\subsection{Finite-range interaction}

In the analysis above, we used a nonzero wave vector $q=2\pi/L$ for the external force in order to satisfy periodic boundary conditions. A side effect of this choice is that the response $1/\epsilon$ remains nonzero even in the absence of pinning, i.e., at $W=0$.
It is straightforward to show that, for $W=0$, the susceptibility with respect to an external potential with wave vector $q$ is given by
\[
\chi_0 = \frac{q^2 + \kappa^2}{2\pi q^2},
\]
provided that the interaction between vortices in Fourier space is defined as
\[
U_\kappa = \frac{2\pi}{q^2 + \kappa^2}.
\]
Our previous results were obtained for the unscreened interaction with $\kappa=0$. In this case,
\begin{equation}
\frac{1}{\epsilon_0}
=
(1 + 2\pi\chi_0)^{-1}
=
0.5,
\label{lowvalue}
\end{equation}
which represents the lower bound for the values of $\langle 1/\epsilon \rangle$ that can be obtained in simulations with nonzero pinning strength $W$.

Introducing a weakly screened interaction with small $\kappa>0$ allows this lower bound to be reduced further, especially for large system sizes $L$:
\begin{equation}
\frac{1}{\epsilon_0(\kappa,L)}
=
(1 + 2\pi\chi_0(\kappa,L))^{-1}
=
\frac{1}{2 + (\kappa L/2\pi)^2}.
\label{lowvalue2}
\end{equation}
Below we present simulation results obtained using such a modification, namely a weakly screened interaction that remains logarithmic at short distances:
\begin{equation}
V(r) = K_0(\kappa r),
\label{Vnum3}
\end{equation}
where $K_0(r)$ is the modified Bessel function. The evaluation of the modified Bessel function was performed using the NAG library routine \texttt{s18acf}.

\begin{figure}[!ht]
\includegraphics[width=1\columnwidth]{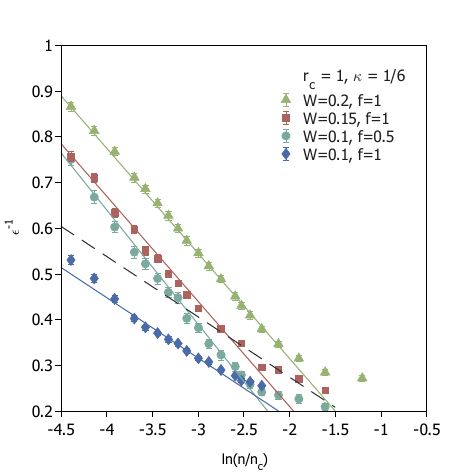}
\centering
\caption{Inverse dielectric function $\langle\varepsilon^{-1}\rangle$ for $L=50$, $\kappa=1/6$, and $r_c=1$, for different disorder strengths $W$, plotted as a function of $\ln(n/n_c)$. The results are fitted by the formula $\langle\varepsilon^{-1}\rangle=- \mathcal{S}\ln(n/n_c)+B$, with $\mathcal{S} =0.23$ and $B=-0.145$ for $W=0.2$ and $f=1$, $\mathcal{S} =0.23$ and $B=-0.25$ for $W=0.15$ and $f=1$, $\mathcal{S} =0.132$ and $B=-0.08$ for $W=0.1$ and $f=1$, and $\mathcal{S} =0.25$ and $B=-0.36$ for $W=0.1$ and $f=0.5$. The dashed line corresponds to the experimental data.}
\label{theoryFig8}
\end{figure}

The simulation results for the inverse dielectric function $\langle \varepsilon^{-1} \rangle$ with $r_c = 1$ and $\kappa = 1/6$ for different disorder strengths, for a system of size $L=50$, are presented in Fig.~\ref{theoryFig8}.

First, we observe that the slope $\mathcal{S}$ increases compared to simulations with an unscreened interaction potential. Second, the saturation value of the inverse dielectric function at large vortex concentrations agrees reasonably well with the analytical prediction
\[
\langle \varepsilon^{-1} \rangle
=
\frac{1}{2 + (\kappa L / 2\pi)^2}
=
0.266.
\]
Indeed, the smallest value obtained in the simulations, $\langle \varepsilon^{-1} \rangle = 0.21$ for $W = 0.1$, is only slightly lower than the predicted value of $0.266$. We attribute this discrepancy to the lattice discretization used in the simulations, which introduces additional pinning effects.

\begin{figure}[!ht]
\includegraphics[width=1\columnwidth]{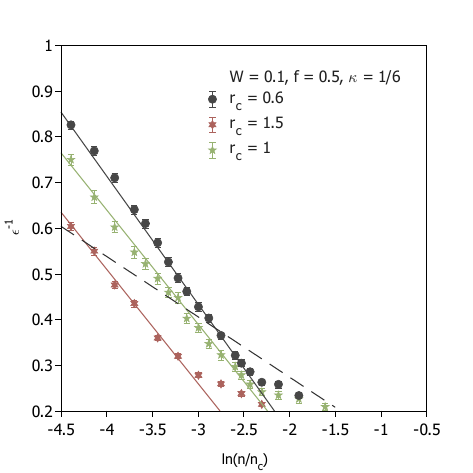}
\centering
\caption{Inverse dielectric function $\langle\varepsilon^{-1}\rangle$ for $L=50$, $\kappa=1/6$, $W=0.1$, and $f=0.5$, for different correlation lengths $r_c$, plotted as a function of $\ln(n/n_c)$. The results are fitted by the formula $\langle\varepsilon^{-1}\rangle=- \mathcal{S}\ln(n/n_c)+B$, with $\mathcal{S} =0.28$ and $B=-0.406$ for $r_c=0.6$, $\mathcal{S} =0.25$ and $B=-0.36$ for $r_c=1$, and $\mathcal{S} =0.25$ and $B=-0.49$ for $r_c=1.5$. The dashed line corresponds to the experimental data.}
\label{theoryFig9}
\end{figure}

In Fig.~\ref{theoryFig9}, we present simulation results for the inverse dielectric function $\langle \varepsilon^{-1} \rangle$ with $L=50$, $\kappa = 1/6$, $W = 0.1$, and $f = 0.5$, for different correlation lengths $r_c$, plotted as a function of $\ln(n/n_c)$.
In contrast to the case of the unscreened interaction, the slope $\mathcal{S}$ exhibits only weak dependence on $r_c$. Specifically, we observe only a slight increase of $\mathcal{S}$ as $r_c$ decreases to $0.6$. However, the resulting slopes remain larger than those observed experimentally.

As follows from the results presented above, introducing the screened interaction potential in Eq.~(\ref{Vnum3}) substantially lowers the saturation value of $\langle \varepsilon^{-1} \rangle$. However, it also increases the slope $\mathcal{S}$ compared to the case of the unscreened logarithmic interaction.

An optimal strategy for simulations in the low-$W$ regime would be to use smaller values of $\kappa \leq 0.05$ together with larger system sizes, such as $L_x = 150$ and $L_y=70$. Under these conditions, the interaction screening would remain sufficiently weak so as not to significantly modify the vortex-glass pinning by disorder, while at the same time keeping the lower bound in Eq.~(\ref{lowvalue2}) sufficiently small.

\subsection{Power-law interaction}

To further clarify the origin of the logarithmic dependence of the inverse dielectric function $1/\epsilon$, we performed analogous simulations using the inter-vortex interaction potential
\[
V(r)=\frac{1}{r}.
\]

Following the same reasoning as discussed around Eqs.~(1) and (2) of the Main Text, one expects in this case
\[
1/\epsilon \propto B^{-0.5}.
\]

\begin{figure}[!ht]
\includegraphics[width=1\columnwidth]{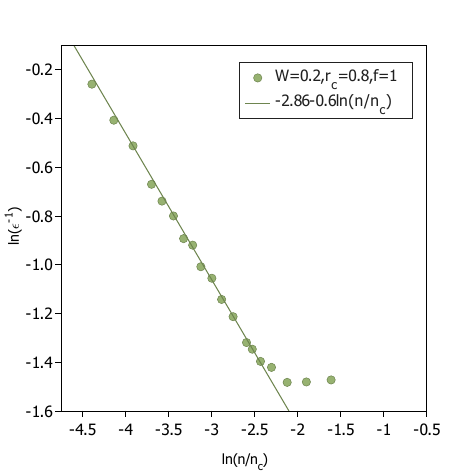}
\centering
\caption{Log-log plot of the inverse dielectric function $\langle\varepsilon^{-1}\rangle$ for $L=50$, $V(r)=1/r$, $W=0.2$, $f=1$, and $r_c=0.8$. The results clearly demonstrate a power-law dependence and are fitted by the formula $\ln(\langle\varepsilon^{-1}\rangle)=-2.86-0.6\ln(n/n_c)$. The saturation observed for $n/n_c \geq 0.1$ is due to the same spurious effect discussed at the beginning of this section.}
\label{theoryFig10}
\end{figure}

The simulation results are presented in Fig.~\ref{theoryFig10}. The figure clearly demonstrates that the inverse dielectric function decreases with vortex density approximately as
\[
\langle\varepsilon^{-1}\rangle\propto (n/n_c)^{-0.6}.
\]

This result further confirms that the logarithmic dependence of $\langle\varepsilon^{-1}\rangle$ observed in our main simulations originates from the logarithmic form of the vortex-vortex interaction.

\section{Analysis of the Hessian spectrum}
In order to find the Hessian it is necessary to calculate the change of energy with respect to small displacements of vortices:
\begin{equation}
 H_{i,j}=\frac{\partial^2E}{\partial \mathbf{r}_i\partial \mathbf{r}_j}
\end{equation}
Here $\mathbf{r}_i$ is the vector coordinate of the i-th vortex. We used here the same energy functional 
with logarithmic repulsion between vortices and a Gaussian random potential, as was used for computations
described above in Sec.I and Sec.II of Supplementary Information.
Since we perform the simulations of vortices using a discrete 2D lattice (with $50 \times 50$ sites), 
the minimal displacement of vortex was restricted by the lattice constant $a$. That leads to a standard finite difference approximation for the spatial derivatives. 
Once the Hessian is calculated, we diagonalize its matrix and compute the density of states $\nu(\lambda) $ of its eigenvalues. We used for this computation a disorder parameter $W = 0.5 g^2$, with a correlation radius
$r_c = 3a$, corresponding to $r_c = 7.5 $ in units of coherence length $\xi_0$.
The results of these calculations are presented in Extended Data Fig.~5.

There are two observations regarding this result. First, $\nu(\lambda)$ for both densities demonstrate a relatively broad maximum for $\lambda$ between 0 and 1, similar to the "boson peak" in glasses, but starting from the lowest $\lambda$.
A second observation is that the Hessian has a small amount of negative eigenvalues. 
This occurs because of the following spurious effect. Vortex configurations  which correspond to any minimum of total energy have some small amount of vortices populating nearest neighbouring sites. Therefore the calculation of the Hessian 
on a lattice requires the calculation of the energy of configurations when two vortices occupy the same site. However, such a configuration was not allowed when we calculated the configuration of minimal energy.
It leads to an ambiguity: the energy of two vortices sitting in the same lattice site is an ill-defined object
for our minimization scheme. To check the effect of this ambiguity we calculated the Hessian spectrum for
several different magnitudes of energy assigned to such "wrong" configurations with overlapping vortices.
It results in some variations within the unphysical "negative tail" of $\nu(\lambda)$, which however
has no effect on the main part of the DoS, $\nu(\lambda >0)$.
Therefore we conclude that  the physical part of the DoS is rather robust and independent of the value of additional two vortex energy.

\section{Independence of superfluid stiffness scaling with mode index}

As the magnetic field increases, higher harmonics of the resonator modes enter the measurement bandwidth, as shown in Fig.~1b. Figure~S12a presents the squared extracted frequencies of the first three modes (the fundamental and first two harmonics) for sample DPres2a on a semilogarithmic scale. Plotting \((f_n(B)/n)^2\) versus magnetic field causes the three curves to collapse, demonstrating that \(f_n(B) \approx n f_1(B)\), as expected for a linear plasmon dispersion relation. This agreement also shows that the extracted \(\Theta(B)\) is frequency independent over our experimentally accessible frequency range.

\begin{figure*}[!ht]
\includegraphics[width=0.9\textwidth]{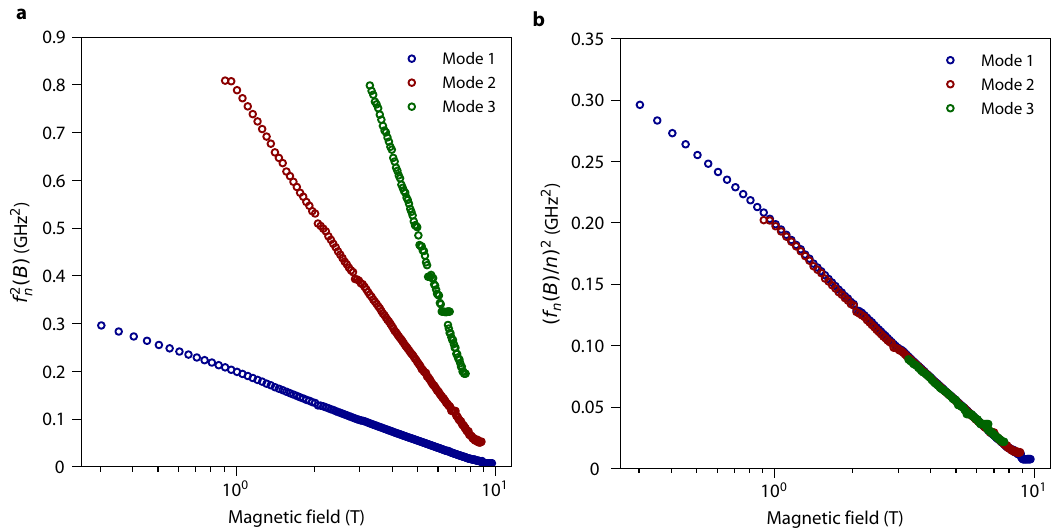}
\centering
\caption{\textbf{Higher harmonics of DPres2a.} \textbf{a} Squared frequency $f^2_n(B)$ of the first three modes ($n = 1, \, 2, \, 3$) of DPres2a versus magnetic field, extracted from Fig.1b. \textbf{b} Same data scaled by the mode index $(f_n(B)/n)^2$. }
\label{Fig-S12}
\end{figure*}




\end{document}